\documentclass[useAMS,usenatbib]{mn2e}

\usepackage{graphicx}
\usepackage{subfigure}
\usepackage{paralist}
\usepackage{amssymb}
\usepackage{bm}
\usepackage{bbding}
\usepackage[fleqn]{amsmath}
\usepackage{multirow}
\usepackage[colorlinks,
            linkcolor=blue, 
            anchorcolor=blue, 
            citecolor=blue, 
            ]{hyperref}
\title[Secular dynamics of stellar spin]{Secular dynamics of stellar spin driven by planets inside Kozai--Lidov resonance}
\author[H. Lei \& Y. Gong]{
	Hanlun Lei$^{1,2}$\thanks{leihl@nju.edu.cn},Yan-Xiang Gong$^{3}$\thanks{yxgong@tsu.edu.cn}\\
	$^{1}$School of Astronomy and Space Science, Nanjing University, Nanjing 210023, China\\
	$^{2}$Key Laboratory of Modern Astronomy and Astrophysics in Ministry of Education, Nanjing University, Nanjing 210023, China\\
	$^{3}$College of Physics and Electronic Engineering, Taishan University, Taian 271000, China
}

\begin{document}

\date{Accepted. Received; in original form}

\pagerange{\pageref{firstpage}--\pageref{lastpage}} \pubyear{2023}

\maketitle
\label{firstpage}

\begin{abstract}
In many exoplanetary systems with `hot Jupiters', it is observed that the spin axes of host stars are highly misaligned to planetary orbital axes. In this study, a possible channel is investigated for producing such a misalignment under a hierarchical three-body system where the evolution of stellar spin is subjected to the gravitational torque induced from the planet inside Kozai--Lidov (KL) resonance. In particular, two special configurations are explored in detail. The first one corresponds to the configuration with planets at KL fixed points, and the second one corresponds to the configurations with planets moving on KL librating cycles. When the planet is located at the KL fixed point, the corresponding Hamiltonian model is of one degree of freedom and there are three branches of libration centres for stellar spin. When the planet is moving on KL cycles, the technique of Poincar\'e section is taken to reveal global structures of stellar spin in phase space. To understand the complex structures, perturbative treatments are adopted to study rotational dynamics. It shows that analytical structures in phase portraits under the resonant model can agree well with numerical structures arising in Poincar\'e sections, showing that the complicated dynamics of stellar spin are governed by the primary resonance under the unperturbed Hamiltonian model in combination with the 2:1 (high-order and/or secondary) spin-orbit resonances.
\end{abstract}


\section{Introduction}
\label{Sect1}

In recent years, more and more exoplanetary systems containing `hot Jupiters' (giant planets with masses $\ge 0.25$ Jupiter's mass and periods $\le 10$ days) are observed to hold high misalignment between the stellar spin axes and planetary orbital axes \citep{albrecht2022stellar,dawson2018origins}. Because of large stellar tidal gravity and radiation fields close to host stars, it is generally believed that hot Jupiters form in regions beyond a few ${\rm AU}$ distance from the host stars and then migrate inward to their current orbits \citep{storch2014chaotic, dawson2018origins}. Moreover, aligned configurations are expected for planet migration in protoplanetary disks (e.g. \citeauthor{bate2010chaotic} \citeyear{bate2010chaotic}). Regarding the puzzle of misaligned `hot Jupiters', \citet{batygin2012primordial} provided a possible explanation that the misalignment exists in the primordial planetary disk relative to the stellar equator. However, if `hot Jupiters' are initially formed in aligned protoplanetary disks, other dynamical channels are required to induce spin-orbit misalignment. In this regard, an overview of three main classes of hot Jupiter origin theory are made by \citet{dawson2018origins}, including in situ formation, gas disk migration and high-eccentricity tidal migration.

To produce retrograde orbits with respect to the total angular momentum, \citet{naoz2011hot} proposed a dynamical channel of secular planet--planet interaction by combining octupole-order gravitational effects with tidal friction. High planetary eccentricities induced by secular interaction stimulates strong planet-star tidal interaction, which can rapidly reduce the orbit energy, leading to inward migration and circularization of planetary orbit and finally forming a retrograde hot Jupiter. This formation channel requires a distant giant planet moving on an eccentric and inclined orbit as a perturber. Due to secular interaction, the angular momentum along the $z$ axis of the inner planet ($H_z$) may change sign, leading to orbit flips. Such a phenomenon is referred to as `eccentric KL effect' \citep{lithwick2011eccentric}. In recent years, varieties of dynamical outcomes and applications of eccentric KL mechanism have been widely explored \citep{lithwick2011eccentric,antognini2015timescales,hamers2021properties,huang2022orbital,lei2022dynamical,lei2022quadrupole,lei2022systematic,li2014,li2014eccentricity,sidorenko2018eccentric,katz2011long,petrovich2015hot}. Now, we know that the eccentric KL effect is due to the apsidal resonance, which is a octupole-order secular resonance under hierarchical planetary systems \citep{sidorenko2018eccentric, lei2022quadrupole,lei2022dynamical,lei2022systematic,huang2022orbital}. Please refer to \citet{naoz2016eccentric} and \citet{shevchenko2016lidov} for an overview about the eccentric KL mechanism and its applications to varieties of astrophysical problems.

In the formation channel proposed in \citet{naoz2011hot}, the stellar spin-orbit coupling is not included, meaning that the stellar equator is always fixed and aligned with the invariant plane of the system. In this sense, the planetary inclination stands for stellar obliquity and thus variation of inclination represents change of stellar obliquity. However, in reality, the central star is an oblate body and it rotates around its spin axis. When the planet moves around the central star on KL cycles (eccentricity and inclination are in coupled oscillations), the rotation-induced stellar quadrupole could produce planet-star interaction torque, forcing the stellar spin axis and planetary orbital axis precess around each other. Based on this fact, \citet{storch2014chaotic} proposed a ``Kozai + tide'' scenario with consideration of stellar spin-orbit coupling. In this scenario, the gravitational interaction between the planet and its oblate host star induced from spin-orbit resonances can cause chaotic evolution of stellar obliquity, showing that stellar spin-orbit misalignment can be produced from aligned configurations. To understand the origin of chaotic behaviours for stellar rotation, \citet{storch2015chaotic} adopted Hamiltonian perturbation theory to deal with a dynamical model where the planets are assumed on periodic KL cycles. In the adiabatic regime (corresponding to regime III in accordance with the classification made in \citeauthor{storch2014chaotic} \citeyear{storch2014chaotic}), they identified a set of secular spin-orbit resonances and showed that the wide-spread chaos in the stellar spin evolution are caused by resonance overlaps. Extending to the non-adiabatic regime (corresponding to regime I in accordance with the classification made in \citeauthor{storch2014chaotic} \citeyear{storch2014chaotic}), \citet{storch2017dynamics} included the effects of short-range forces and tidal dissipation and categorised different paths to spin-orbit misalignment. They pointed out two spin-orbit evolution paths can lead to retrograde configurations. Considering the influence of the octupole-level effects, \citet{anderson2017eccentricity} derived the required condition for producing spin-orbit misalignment in the inner binary. 

In the previous studies \citep{storch2014chaotic, storch2015chaotic, storch2017dynamics}, the authors assumed that the planets are placed on fixed KL circulating cycles (close to KL separatrix), which are outside the KL resonance. Thus, it is unclear about the influence of different-scale KL cycles upon the secular dynamics of stellar spin. In addition, a more formal canonical perturbation theory is absent in their series of studies. Inspired by these considerations, we revisit the stellar spin dynamics in this work under the configurations with planets inside KL resonance. In particular, two special configurations are considered: the first one with planets located at the KL fixed point and the second one with planets moving on KL librating cycles. In the configuration with planets at the KL fixed point, the resulting Hamiltonian determines a one-degree-of-freedom dynamical model and distributions of the so-called `Cassini's states' (i.e., equilibrium points under the 1 DOF Hamilotnian model) are produced. In the configuration with planets moving on KL librating cycles, numerical technique of Poincar\'e section is adopted to obtain the global structures of stellar spin in phase space, and then theory of perturbative treatment is taken to understand the complex structures arising in Poincar\'e sections. It shows that analytical structures in phase portraits under the resonant model can agree well with numerical structures arising in Poincar\'e sections, making clear the dynamical mechanism governing the spin structures.

The remaining part of this work is organised as follows. In Section \ref{Sect2}, the Hamiltonian functions governing stellar spin evolution and planet's KL oscillations are briefly introduced. In Section \ref{Sect3}, secular dynamics of stellar spin are investigated under a special configuration with planets locating at the KL centre. Section \ref{Sect4} studies stellar spin dynamics under the configurations with planets moving on KL cycles by taking advantage of numerical approach (Poincar\'e sections) and analytical approach (perturbative treatments). Conclusions are summarised in Section \ref{Sect5}.

\section{Hamiltonian model}
\label{Sect2}

\begin{figure}
	\centering
	\includegraphics[width=\columnwidth]{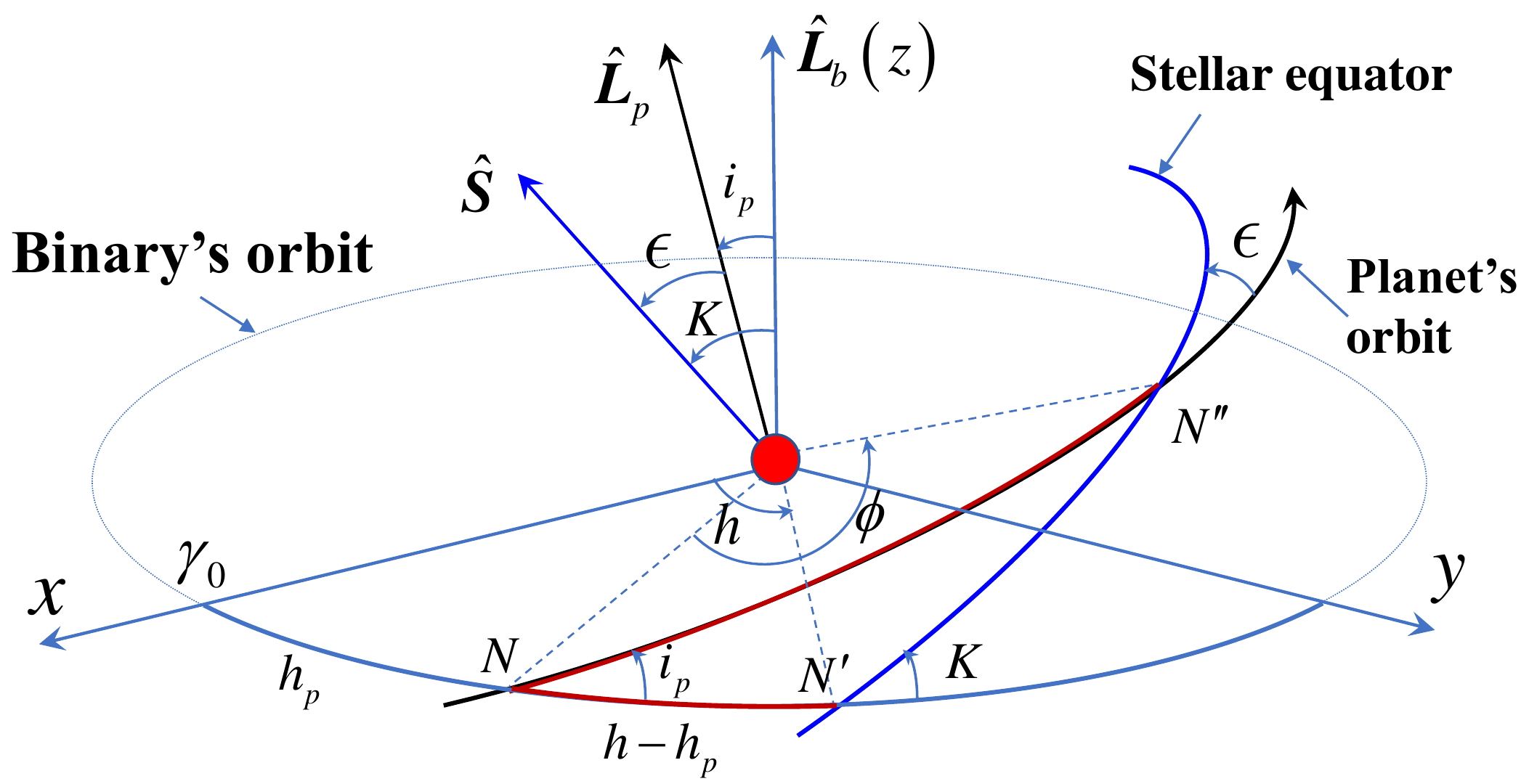}
	\caption{Definition of the variables used in this work and relative configuration of three fundamental planes: the invariant plane of binary system, orbital plane of the planet around the central star and the stellar equator. Normal directions of these planes are denoted by unitary vectors ${\hat {\bm L}}_b$, ${\hat {\bm L}}_p$ and ${\hat {\bm S}}$. The ascending nodes are denoted by $N$, $N'$ and $N''$, and the longitudes of ascending nodes are $h_p$, $h$ and $\phi$. The orbital inclination of planet relative to the binary's orbit is $i_p$, the absolute obliquity of stellar equator relative to the binary's orbit is $K$ and the relative obliquity of stellar equator with respect to planet's orbit is $\epsilon$. Here, $\epsilon$ can be used to measure the `orbital inclination' of planets relative to stellar equator. The point $\gamma_0$ denotes the pericentre of binary's orbit.}
	\label{Fig1}
\end{figure}

In this study, we consider a hierarchical three-body system, consisting of an oblate star with mass $M_*$, an inner giant planet with mass $m_p$ and a distant binary companion with mass $m_b$. In practical simulations, we take $M_* = m_b \gg m_p$, as adopted by \citet{storch2015chaotic} and \citet{storch2017dynamics} in their studies. The planet moves around the central star in the gravitational field generated by the stellar binary. For simplicity, the following two assumptions are made about the dynamical system \citep{storch2014chaotic, storch2015chaotic, storch2017dynamics}: (a) the evolution of stellar obliquity is dominated by the gravitational torque induced from the planet, and (b) the orbital motions of the planet and perturber around the central star are not influenced by stellar rotation (i.e., the back-reaction from the stellar rotation to orbital evolution is ignored). The second assumption indicates that the orbital evolution of the considered system is decoupled from stellar rotation. Thus, the orbital dynamics can be studied separately. Relaxing the second assumption to a coupled case will be performed in future.

Figure \ref{Fig1} shows the relative geometry of the orbital plane of binary, orbital plane of the planet and stellar equator. ${\bm L}_b$ stands for the angular momentum vector of the binary's orbit, ${\bm L}_p$ is the angular momentum vector of the planet's orbit and $\bm S$ represents the stellar spin axis vector. Their magnitudes are ${L}_b$, ${L}_p$ and $S$ and their unitary vectors are denoted by ${\hat {\bm L}}_b$, ${\hat {\bm L}}_p$ and ${\hat {\bm S}}$. For the considered system, it holds ${L}_b \gg L_p \gg S$. Thus, it is reasonable to  assume the orbital plane of binary as the invariant plane of system. Based on the invariant plane, an inertial coordinate system (named $O$--$xyz$) is introduced: the $x$-axis directs from the central star towards  binary's pericentre $\gamma_0$, the $z$-axis is along the angular momentum vector of the binary and the $y$-axis completes a right-handed system (see Fig. \ref{Fig1}). The ascending nodes of three fundamental planes are denoted by $N$, $N'$ and $N''$, and the corresponding longitudes of ascending nodes are $h$, $h_p$ and $\phi$ (see Fig. \ref{Fig1} for detailed definitions). The relative angle between ${\bm L}_p$ and ${\bm L}_b$ corresponds to the inclination of planet's orbit ($i_p$), the angle between ${\bm L}_b$ and ${\bm S}$ is the absolute obliquity of the stellar equator relative to the invariant plane ($K$), and the angle between ${\bm L}_p$ and ${\bm S}$ is the relative obliquity of stellar equator with respect to its orbit around planet ($\epsilon$). According to the above definitions, we have the following relations:
\begin{equation*}
	{{\hat {\bm L}}_p} \cdot {{\hat {\bm L}}_b} = \cos {i_p},\quad	\hat {\bm S} \cdot {{\hat {\bm L}}_b} = \cos K,\quad	\hat {\bm S} \cdot {{\hat {\bm L}}_p} = \cos \epsilon.
\end{equation*}
Among these variables, $(h,\cos{K})$, $(h_p, \cos{i_p})$ and $(\phi, \cos{\epsilon})$ are three pairs of conjugate variables. For simplicity, we denote $p=\cos{\epsilon}$. The transformations between these sets of conjugate variables can be realised by
\begin{equation*}
	\begin{aligned}
		\sin \epsilon \sin \phi  =& \sin K\sin \left( {h - {h_p}} \right),\\
		- \sin \epsilon \cos \phi  =& \cos K\sin {i_p} - \sin K\cos {i_p}\cos \left( {h - {h_p}} \right),\\
		\cos \epsilon  =& \cos K\cos {i_p} + \sin K\sin {i_p}\cos \left( {h - {h_p}} \right),
	\end{aligned}
\end{equation*}
and
\begin{equation*}
	\begin{aligned}
		\sin K\sin \left( {h - {h_p}} \right) =& \sin \epsilon \sin \phi,\\
		- \sin K\cos \left( {h - {h_p}} \right) =&  - \cos \epsilon \sin {i_p} - \sin \epsilon \cos {i_p}\cos \phi,\\
		\cos K =& \cos \epsilon \cos {i_p} - \sin \epsilon \sin {i_p}\cos \phi.
	\end{aligned}
\end{equation*}
Under the coordinate system $O$--$xyz$, the unitary vector ${\hat {\bm L}}_b$ is along the $z$-axis, the unitary vectors ${\hat {\bm L}}_p$ and ${\hat {\bm S}}$ are given by
\begin{equation*}
	{\hat {\bm L}_p} = \left[ {\begin{array}{*{20}{c}}
			{\sin {i_p}\sin {h_p}}\\
			{ - \sin {i_p}\cos {h_p}}\\
			{\cos {i_p}}
	\end{array}} \right],\quad \hat {\bm S} = \left[ {\begin{array}{*{20}{c}}
			{\sin K\sin h}\\
			{ - \sin K\cos h}\\
			{\cos K}
	\end{array}} \right].
\end{equation*}

\subsection{Kozai--Lidov dynamics}
\label{Sect2-1}

As assumed above, orbital evolution of planet is decoupled from stellar rotation, thus the orbital dynamics can be described separately. Since $m_p$ is much smaller than $M_*$ and $m_b$, the gravitational influence coming from the planet upon the binary can be ignored. Thus, the binary moves around their barycentre in Keplerian orbits, and the planet moves around the central star in a perturbed Keplerian orbit under perturbation from the distant perturber. In the coordinate system $O$--$xyz$, the perturber's orbit is characterised by the semi-major axis $a_b$ and eccentricity $e_b$, and the planet's orbit is characterised by the semi-major axis $a_p$, eccentricity $e_p$, inclination $i_p$, longitude of ascending node $\Omega_p$, argument of pericentre $\omega_p$ and mean anomaly $M_p$. In hierarchical configurations, $a_p$ is much smaller than $a_b$, thus the orbital dynamics of the planet can be approximated by the quadruple-order Hamiltonian.

For convenience, we introduce Delaunay variables as follows \citep*{morbidelli2002modern}:
\begin{equation*}
	\begin{aligned}
		{l_p} =&\; {M_p},\quad {L_p} = \sqrt {\mu {a_p}}, \\
		{g_p} =&\; {\omega _p},\quad {G_p} = {L_p}\sqrt {1 - e_p^2}, \\
		{h_p} =&\; {\Omega _p},\quad {H_p} = {G_p}\cos {i_p},
	\end{aligned}
\end{equation*}
where the gravitational parameter is $\mu  = {\cal G}\left( {{M_*} + {m_p}} \right)$. Under the influence of generality relativity (GR) effect, the quadrupole-order Hamiltonian, averaged over the orbital periods of planet and perturber, can be written as \citep*{kozai1962secular,wu2003planet,liu2015suppression,naoz2016eccentric}
\begin{equation}\label{Eq1}
	\begin{aligned}
		{{\cal H}_{{\rm{KL}}}} = \; &- \left( {5 - 3\frac{{G_p^2}}{{L_p^2}}} \right)\left( {3\frac{{H_p^2}}{{G_p^2}} - 1} \right)\\
		&- 15\left( {1 - \frac{{G_p^2}}{{L_p^2}}} \right)\left( {1 - \frac{{H_p^2}}{{G_p^2}}} \right)\cos 2{g_p} - \frac{{3{\mu ^4}\beta }}{{{{\cal C}_0}L_p^3{G_p}{c^2}}}
	\end{aligned}
\end{equation}
where $c$ is the speed of light and the coefficient ${\cal C}_0$ is given by
\begin{equation*}
	{{\cal C}_0} = \frac{1}{{16}}\frac{{{\cal G}{m_b}}}{{{a_b}}}\beta {\left( {\frac{{{a_p}}}{{{a_b}}}} \right)^2}\frac{1}{{{{\left( {1 - e_b^2} \right)}^{3/2}}}}
\end{equation*} 
with $\beta$ being the reduced mass $\beta  = \frac{{{M_*}{m_p}}}{{{M_*} + {m_p}}}$. From Lagrange planetary equations \citep{murray1999solar}, it is possible to derive the time derivatives of $i_p$ and $h_p$ as follows:
\begin{equation}\label{Eq2}
	\begin{aligned}
		\frac{{{\rm d}{i_p}}}{{{\rm d}t}} =\;&  - \frac{{15}}{{{G_p}}}e_p^2\sin 2{i_p}\sin 2{g_p},\\
		\frac{{{\rm d}{h_p}}}{{{\rm d}t}} =\;& \frac{6}{{{G_p}}}\left[ { - \left( {2 + 3e_p^2} \right)\cos {i_p} + 5e_p^2\cos {i_p}\cos 2{g_p}} \right].
	\end{aligned}
\end{equation}
Equation (\ref{Eq2}) determines the evolution of the unitary vector $\hat {\bm L}_p$ (normal direction of planet's orbit) and it will be used in the formulation of stellar spin Hamiltonian model. 

The last term arising in equation (\ref{Eq1}) stands for the GR effect. It is known that the GR effect usually tends to reduce the maximum eccentricity $e_{\max}$ reached by a KL cycle but does not change the dynamical structures \citep{storch2017dynamics}. If the GR effect is ignored, it becomes the well-known Hamiltonian for studying Kozai--Lidov (KL) resonance \citep{kozai1962secular,naoz2016eccentric}. The dynamical model represented by equation (\ref{Eq1}) is of one degree of freedom, depending on the motion integral $H_p = G_p \cos{i_p} = {L_p}\sqrt {1 - e_p^2} \cos{i_p}$. In the long-term evolution, the semi-major axis as well as the $z$-component of angular momentum $H_p$ remains unchanged \citep{naoz2016eccentric}. Usually, the motion integral $H_p$ can be specified by the maximum inclination $i_{\max}$ by means of $H_p = L_p \cos{i_{\max}}$ \citep*{kozai1962secular}. It is known that, when $i_{\max}$ is greater than a critical inclination, KL resonance can take place. For the dynamical model without GR effect, the critical inclination is equal to $39.2^{\circ}$ or $140.8^{\circ}$ \citep*{kozai1962secular}. However, when the GR effect is taken into account, the critical inclination may change, as shown later. 

\begin{figure*}
	\centering
	\includegraphics[width=\columnwidth]{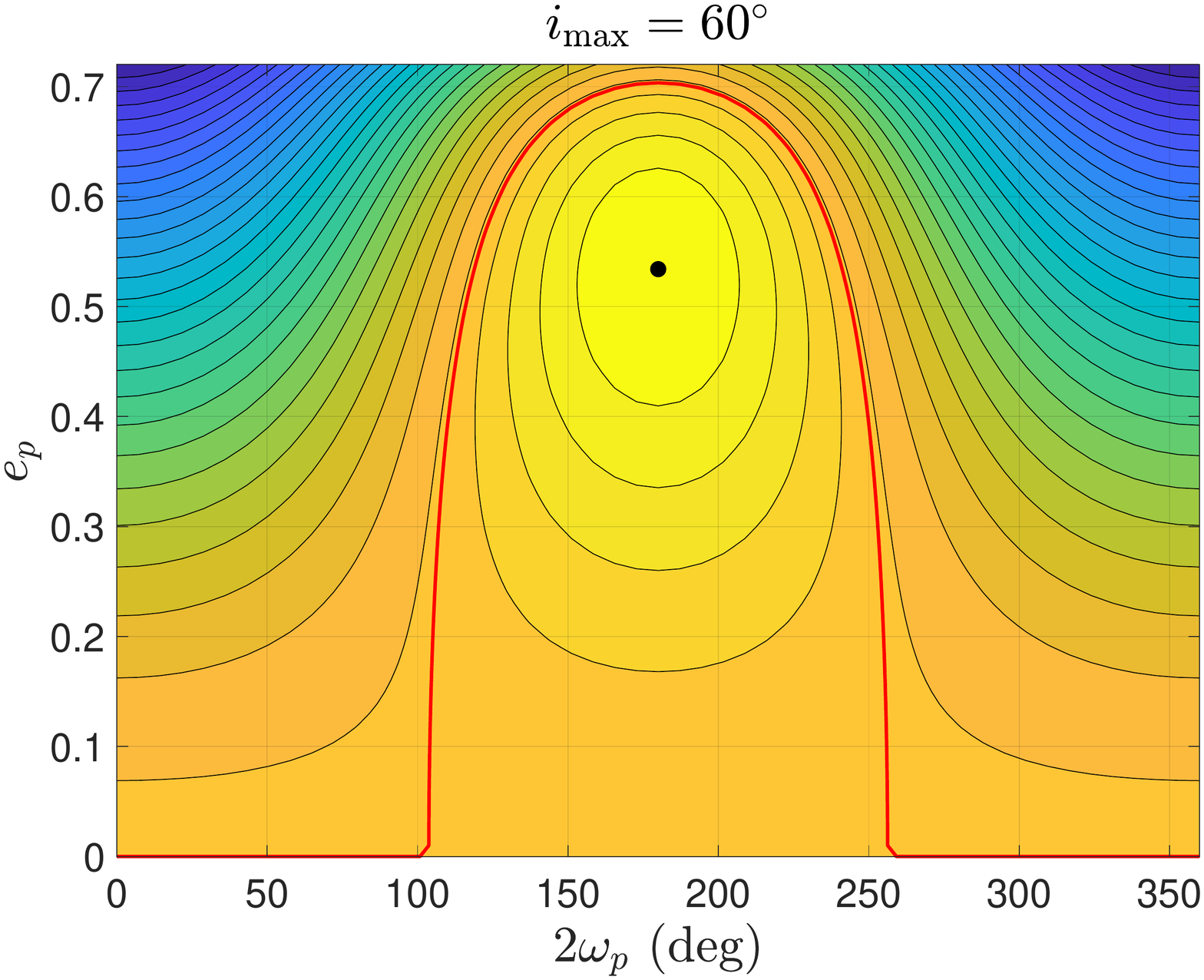}
	\includegraphics[width=\columnwidth]{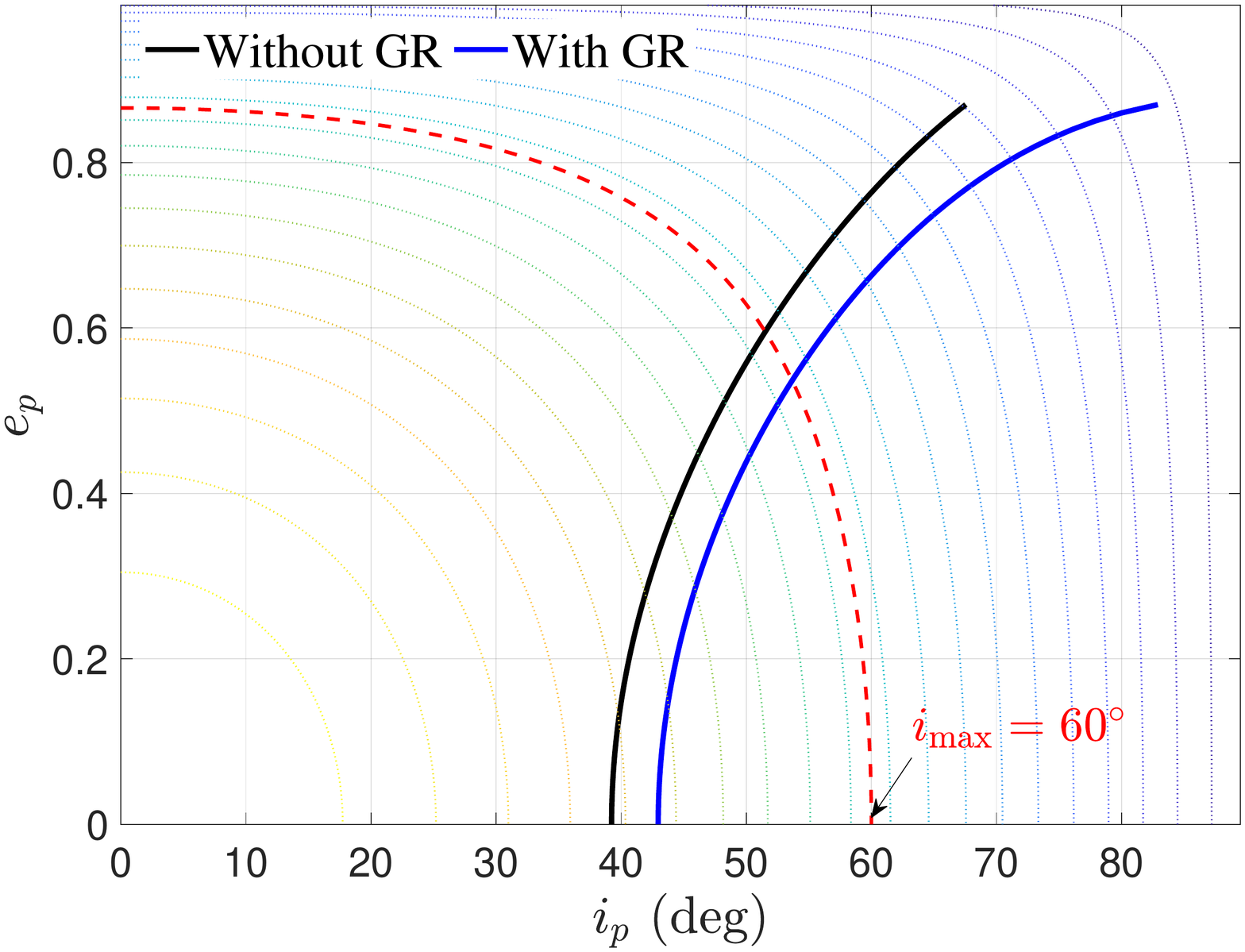}
	\caption{Dynamical structure at the quadrupole-level approximation with GR effect when the maximum inclination $i_{\max}$ is equal to $60^{\circ}$ (\emph{left panel}) and the distribution of KL centres in the $(i_p, e_p)$ panel for the model with and without GR effects (\emph{right panel}). In the left panel, the dynamical separatrix, dividing the phase space into regions of circulation and libration, is shown by a red line, and the KL centre is marked by a black dot. In the right panel, the level curves of the motion integral $H_p =G_p\cos{i_p}$ are presented as background and the level curve corresponding to $i_{\max}=60^{\circ}$ is shown by a red dashed line.}
	\label{Fig2}
\end{figure*}

With the GR effect, the location of Kozai--Lidov centre is determined by the following equality:
\begin{equation}\label{Eq2-1}
	{\left( {1 - e_p^2} \right)^{3/2}} - \frac{5}{3}{\left( {1 - e_p^2} \right)^{1/2}}{\cos ^2}{i_p} = \frac{1}{{12}}\frac{{{\mu ^2}\beta }}{{{{\cal C}_0}a_p^2{c^2}}},
\end{equation}
which determines the critical inclination $i_c$ (at $e_p = 0$) by
\begin{equation*}
	{\cos ^2}{i_c} = \frac{3}{5}\left( {1 - \frac{1}{{12}}\frac{{{\mu ^2}\beta }}{{{{\cal C}_0}a_p^2{c^2}}}} \right).
\end{equation*}
Obviously, the value of $i_c$ is dependent on the planet's semimajor axis $a_p$ as well as mass $m_p$. 

Without the GR effect, the equality for determining the location of Kozai--Lidov centre becomes
\begin{equation*}
	e_p^2 + \frac{5}{3}{\cos ^2}{i_p} = 1,
\end{equation*}
which determines the critical inclination $i_c$ (at $e_p = 0$) by
\begin{equation*}
	{\cos ^2}{i_c} = \frac{3}{5} \Rightarrow {i_c} = 39.2^{\circ} \;{\rm or}\; 140.8^{\circ}.
\end{equation*}
It should be mentioned that the angular coordinate of KL centre is at $2g_p = \pi$.

Dynamical structure of quadrupole-level Hamiltonian with GR effect is shown in the left panel of Fig. \ref{Fig2}. The parameters of dynamical model are to be described in Section \ref{Sect2-3}. When the maximum inclination is at $i_{\max} = 60^{\circ}$, the KL centre is located at $(2\omega_p = 0, e_ p = 0.534)$. In the right panel of Fig. \ref{Fig2}, KL centres under the dynamical models with and without GR effect are shown in the $(i_p, e_p)$ plane. For convenience, the level cures of the motion integral $H_p$ are presented as background and the specific curve corresponding to $i_{\max}=60^{\circ}$ is shown by a red dashed line. It is observed that KL centres with GR effect are located on the right hand of the ones without GR effect, showing that KL centres with GR effect hold higher inclinations at a given eccentricity. 

The KL cycles under the quadrupole-level Hamiltonian are periodic and their analytical expressions can be provided in terms of elliptic functions \citep{kinoshita2007general}. On the other hand, the dynamical model shown by equation (\ref{Eq1}) is of one degree of freedom, thus the KL cycles under this model correspond to the level curves of Hamiltonian in the phase space (see the left panel of Fig. \ref{Fig2}). The dynamical separatrix, shown by a red line, can divide the entire phase space into regions of libration and circulation. Those trajectories inside the island of libration are referred to as KL librating cycles and the ones outside the island are called KL circulating cycles.

\subsection{Hamiltonian of stellar spin}
\label{Sect2-2}

Because we are interested in the evolution of stellar obliquity $\epsilon$ driven by the planet, it is convenient to formulate the Hamiltonian governing the dynamics of stellar spin in the rotating frame in which the $x$-axis is directed from the star towards the ascending node $N$, the $z$-axis is along the vector $\hat{\bm L}_p$ and the $y$-axis is chosen to complete a right-handed coordinate system. In this rotating frame, the normalised Hamiltonian can be written as \citep{storch2015chaotic}
\begin{equation}\label{Eq3}
	\begin{aligned}
		{\cal H} =  - \frac{1}{{{{\cal C}_0}}}\frac{{3{\cal G}{m_p}\left( {{I_3} - {I_1}} \right)}}{{4a_p^3{{\left( {1 - e_p^2} \right)}^{3/2}}}}\frac{{{{\cos }^2}\epsilon }}{{{S^{\rm{*}}}}} - \frac{{{\bm R} \cdot {{\bm S}^{\rm{*}}}}}{{{S^{\rm{*}}}}},
	\end{aligned}
\end{equation}
where $\bm S^*$ is the scaled stellar spin axis vector ($\bm S^* = {\bm S}/\beta$) and its magnitude is denoted by $S^*$. Here, $I_3$ and $I_1$ stand for the principal moments of inertia of the central star, defined by
\begin{equation*}
	{I_3} - {I_1} = {k_q}{M_*}R_*^2\hat \Omega _*^2,
\end{equation*}
where ${\hat \Omega _*} = {{\Omega }_*}/\sqrt {\frac{{{\cal G}{M_*}}}{{R_*^3}}}$ is the dimensionless stellar spin frequency, $R_*$ is the stellar radius and $S = {k_*}{M_*}R_*^2 {\Omega}_*$. For a solar-type star, it holds $k_q = 0.05$ and $k_* = 0.1$ \citep{storch2015chaotic}. 

In equation (\ref{Eq3}), $\bm R$ is the rotational vector of planet's orbit relative to the invariant plane, given by \citep{kinoshita1993motion,storch2015chaotic},
\begin{equation}\label{Eq4}
	{\bm R} = \frac{{{\rm d}{h_p}}}{{{\rm d}t}}{\hat {\bm L}_b} + \frac{{{\rm d}{i_p}}}{{{\rm d}t}}\left( {\frac{{{{\hat {\bm L}}_b} \times {{\hat {\bm L}}_p}}}{{\sin {i_p}}}} \right) = \left[ {\begin{array}{*{20}{l}}
			{\frac{{{\rm d}{i_p}}}{{{\rm d}t}}}\vspace{1ex}\\
			{\frac{{{\rm d}{h_p}}}{{{\rm d}t}}\sin {i_p}}\vspace{1ex}\\
			{\frac{{{\rm d}{h_p}}}{{{\rm d}t}}\cos {i_p}}
	\end{array}} \right],
\end{equation}
and $\bm S^*$ is the rotational angular momentum vector of the star measured in the rotating frame by
\begin{equation}\label{Eq5}
	{\bm S^*} = {S^*}\left[ {\begin{array}{*{20}{r}}
			{\sin \epsilon \sin \phi }\\
			{ - \sin \epsilon \cos \phi }\\
			{\cos \epsilon }
	\end{array}} \right].
\end{equation}
Replacing equations (\ref{Eq4}) and (\ref{Eq5}) in equation (\ref{Eq3}), we can get the Hamiltonian as follows:
\begin{equation}\label{Eq6}
	\begin{aligned}
		{\cal H} =&  - \frac{{3{{\cal C}_1}}}{{2{{\cal C}_0}{S^*}G_p^3}}{\cos ^2}\epsilon  - \sin \epsilon \sin \phi \frac{{{\rm d}{i_p}}}{{{\rm d}t}}\\
		&- \left( {\cos \epsilon \cos {i_p} - \sin \epsilon \sin {i_p}\cos \phi } \right)\frac{{{\rm d}{h_p}}}{{{\rm d}t}},
	\end{aligned}
\end{equation}
where the coefficient ${\cal C}_1$ is given by
\begin{equation*}
	{{\cal C}_1} = \frac{{{\cal G}{m_p}{\mu ^{3/2}}\left( {{I_3} - {I_1}} \right)}}{{2a_p^{3/2}}}.
\end{equation*}
Substituting $\frac{{{\rm d}{i_p}}}{{{\rm d}t}}$ and $\frac{{{\rm d}{h_p}}}{{{\rm d}t}}$ given by equation (\ref{Eq2}) into equation (\ref{Eq6}), one can obtain the explicit expression of Hamiltonian for stellar spin in terms of conjugate variables $(\phi, p = \cos{\epsilon})$ as follows:
\begin{equation}\label{Eq7}
	\begin{aligned}
		{\cal H} =&  - \frac{{3{{\cal C}_1}}}{{2{{\cal C}_0}{S^*}G_p^3}}{p^2} + \frac{{6H_p^2}}{{L_p^2G_p^3}}\left[ {5L_p^2 - 3G_p^2}- 5\left( {L_p^2 - G_p^2} \right) \right.\\
		&\left. \times{\cos 2{g_p}} \right] p+ \frac{{6{H_p}}}{{L_p^2G_p^3}}\sqrt {\left( {1 - {p^2}} \right)\left( {G_p^2 - H_p^2} \right)} \\
		&\times \left[ {5\left( {L_p^2 - G_p^2} \right)\cos \left( {2{g_p} - \phi } \right) - \left( {5L_p^2 - 3G_p^2} \right)\cos \phi } \right],
	\end{aligned}
\end{equation}
where $(g_p,G_p)$ are known periodic functions of time (with the same period of KL oscillation), determined by the KL Hamiltonian. As the Hamiltonian (\ref{Eq7}) is time dependent, it determines a non-autonomous dynamical model and it is no longer a constant during the long-term evolution. It should be mentioned that an equivalent expression of spin Hamiltonian (without GR effect) can be found in \citet{storch2015chaotic} under a different notation system (see equation (30) in their work). 

The equations of motion can be derived from Hamiltonian canonical relations:
\begin{equation}\label{Eq8}
	\begin{aligned}
		\dot \phi  =& \frac{{\partial {\cal H}}}{{\partial p}},\quad \dot p =  - \frac{{\partial {\cal H}}}{{\partial \phi }},\\
		\dot g_p  =& \frac{{\partial {\cal H_{\rm KL}}}}{{\partial G_p}},\quad \dot G_p =  - \frac{{\partial {\cal H_{\rm KL}}}}{{\partial g_p}},
	\end{aligned}
\end{equation}
where $\cal H_{\rm KL}$ is given by equation (\ref{Eq1}) and $\cal H$ is given by equation (\ref{Eq7}). In the long-term spin-orbit evolution, $L_p$ and $H_p$ remain unchanged (because $l_p$ and $h_p$ are cyclic variables). 

It should be once again mentioned that the evolution of $(g_p,G_p)$ is decoupled from the stellar rotation state $(\phi,p)$ while the evolution of $(\phi,p)$ is dependent on the orbital state $(g_p,G_p)$. The time histories $g_p (t)$ and $G_p (t)$ governed by  $\cal H_{\rm KL}$ are called KL cycles, which are periodic functions.

\subsection{Numerical integration}
\label{Sect2-3}
Unless otherwise stated, the following representative parameters of system are adopted for simulations performed in the entire work:
\begin{equation*}
	\begin{aligned}
		{{\hat \Omega }_*} &= 0.05,\quad k_q=0.05,\quad k_* = 0.1,\quad R_* = 1.0\,R_{\sun},\\
		M_* &= 1.0\,M_{\sun},\quad m_b = 1.0\, M_{\sun},\quad m_p = 5.0\,M_{J},\\
		a_p &= 1.0\,{\rm AU},\quad a_b = 200\,{\rm AU},\quad e_b = 0,
	\end{aligned}
\end{equation*}
where $M_{\sun}$ is the mass of the Sun, $R_{\sun}$ is the radius of the Sun and $M_{J}$ is the mass of the Jupiter. Because of the small semi-major axis ratio $\alpha = 1/200$, the octupole-order and higher-order influences are significantly weak, compared to the leading-order term. Moreover, in the configuration with the binary companion's orbit being circular ($e_b = 0$), the octupole-order effect vanishes. Thus, the quadrupole-level Hamiltonian is adequate to describe the KL cycles.

For convenience of computation, we take the mass of the Sun as the mass unit and the mean distance between the Sun and Earth ($\rm AU$) as the unit of length. The time unit is taken to make the orbital period of the Earth around the Sun be $2\pi$.  Under the system of dimensionless units, the universal gravitation constant $\cal G$ and the mean motion of the Earth are unitary, and the coefficients ${\cal C}_0$ and ${\cal C}_1$ are
\begin{equation*}
	{\cal C}_0 = 3.711935\times 10^{-11},\quad {\cal C}_1 = 6.499119\times 10^{-12}.
\end{equation*}
With these parameters, the critical inclination $i_c$ with consideration of GR effect is
\begin{equation*}
	i_c = 42.9^{\circ}\;{\rm or}\;137.1^{\circ}.
\end{equation*}
Thus, with consideration of the GR effect, Kozai--Lidov resonance happens in the interval of $i_{\max} \in \left[42.9^{\circ}, 137.1^{\circ}\right]$, which is narrower than the conventional interval of $i_{\max} \in \left[39.2^{\circ}, 140.8^{\circ}\right]$.

\begin{figure*}
	\centering
	\includegraphics[width=\columnwidth]{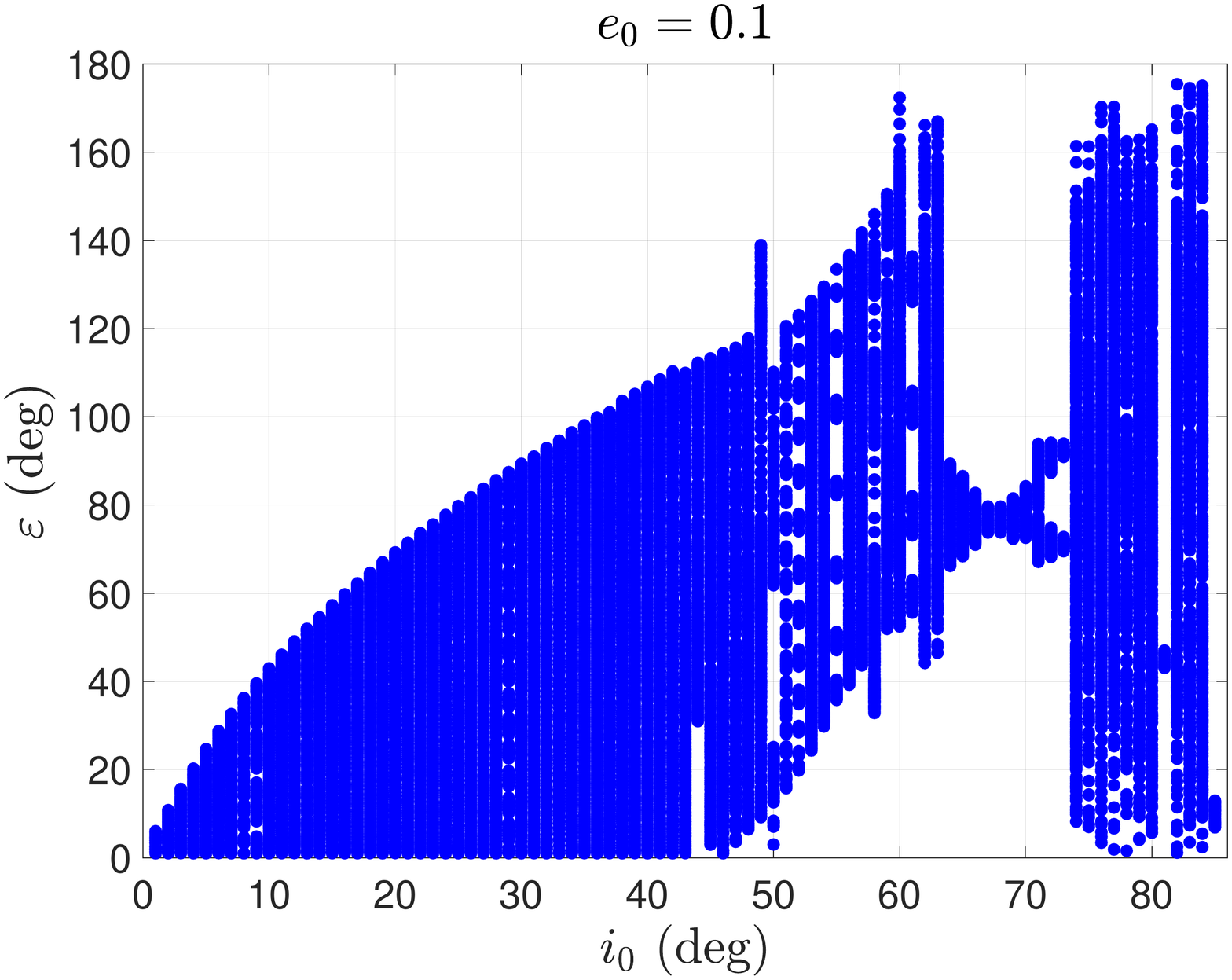}
	\includegraphics[width=\columnwidth]{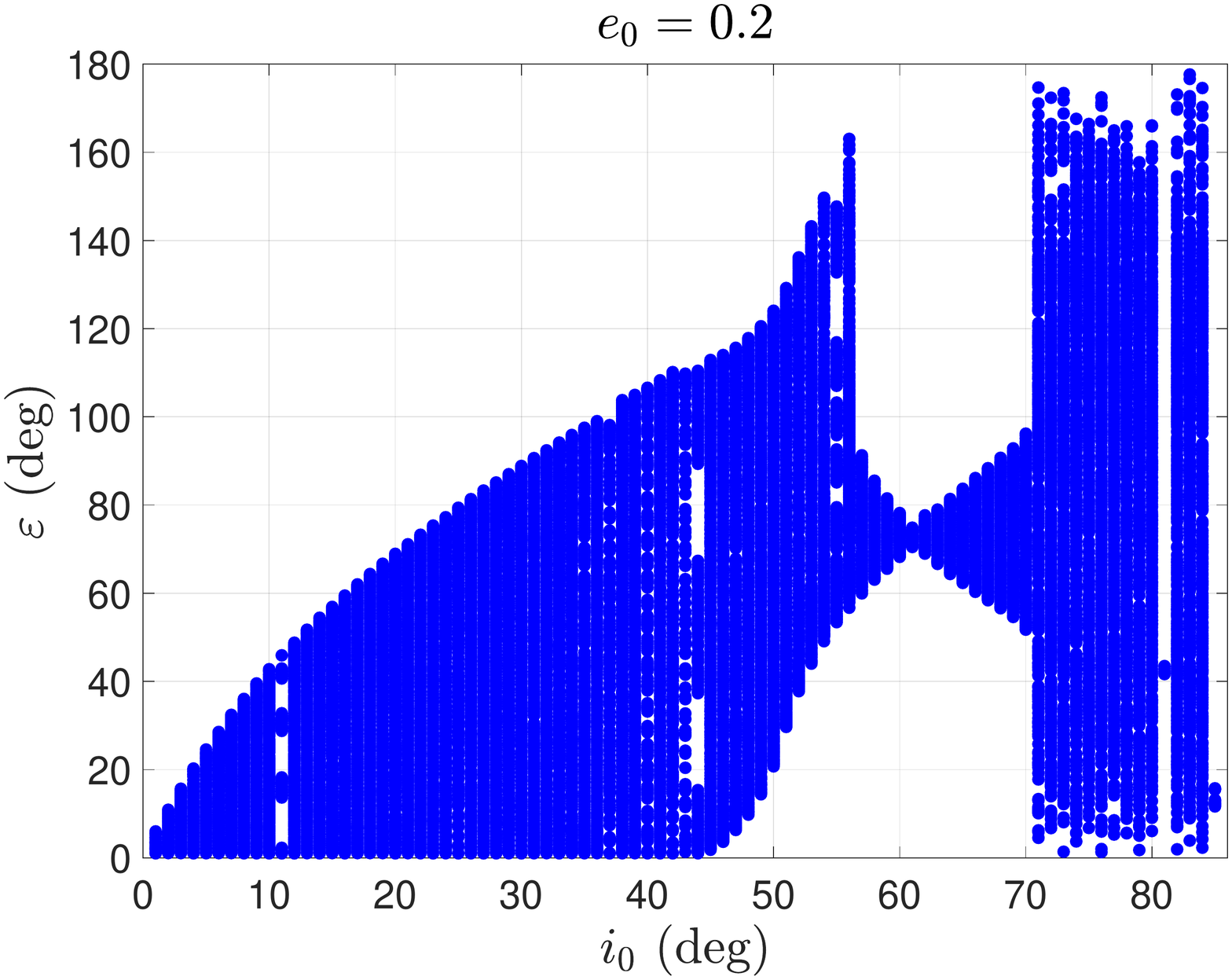}
	\caption{`Bifurcation' diagrams of stellar spin-orbit misalignment angle as a function of the initial inclination $i_0$ for the cases of $e_0 = 0.1$ (\emph{left panel}), $e_0 = 0.2$ (\emph{right panel}). For producing the `bifurcation' diagrams, the equations of motion represented by equation (\ref{Eq8}) are numerically integrated over 500 KL periods with initial conditions $\omega_0 = \pi/2$, $\epsilon_0 = 1.0^{\circ}$ and $\phi_0 = 0^{\circ}$ and the states of stellar spin are recorded every time the eccentricity reaches a maximum.}
	\label{Fig3}
\end{figure*}

To see the variation of stellar obliquity, we numerically integrate the equations of motion represented by equation (\ref{Eq8}) over 500 KL periods and record the state of stellar spin every time the eccentricity reaches a maximum (this process is similar to producing Poincar\'e section defined by $g_p = \pi/2$ and ${\dot g}_p >0$). The initial conditions are taken as $\omega_0 = \pi/2$, $\epsilon_0 = 1.0^{\circ}$ (a nearly aligned configuration) and $\phi_0 = 0^{\circ}$. Figure \ref{Fig3} shows the distribution of stellar obliquity as a function of the initial inclination $i_0$ for the cases of $e_0 = 0.1$ (in the left panel) and $e_0 =0.2$ (in the right panel). \citet{storch2014chaotic} and \citet{storch2015chaotic} referred to such a kind of distribution as `bifurcation' diagrams. From Fig. \ref{Fig3}, we can observe: (a) `bifurcation' diagrams are changed with different initial eccentricities, (b) the misalignment angle can reach higher than $90^{\circ}$ from an aligned configuration when the initial inclination is larger than $\sim$$30^{\circ}$, (c) multiple periodic islands or spin-orbit resonances can be observed from the `bifurcation' diagrams, (c) the evolution of stellar spin behaves regular when the initial inclination $i_0$ is smaller than $\sim$$50^{\circ}$ and, within this interval, the maximum obliquity increases with initial inclination, and (d) both chaotic and regular behaviours of stellar spin can be found when the initial inclination is greater than $\sim$$50^{\circ}$, meaning that in this interval the evolution of stellar obliquity is complex. It is mentioned that chaotic spin-orbit resonances are popular in the solar system, such as Saturn's satellite Hyperion \citep{wisdom1984chaotic}, terrestrial planets \citep{laskar1993chaotic}, etc. 

To understand the complex dynamics of stellar spin driven by planets, in the following sections, we concentrate on two cases under the GR effect: (i) the planet is located at the KL fixed point and (ii) the planet is moving on KL librating cycles. The first case can approximate stellar spin dynamics within the configurations deeply inside KL resonance. The second case can be used to approximate stellar spin dynamics within more general configurations inside KL resonances. 

\section{Spin dynamics at KL fixed point}
\label{Sect3}

The location of KL fixed point $(e_k, i_k, 2g_k = \pi)$ can be determined by equation (\ref{Eq2-1}). Thus, the distribution of KL fixed point is a function of $i_{\max}$ (please see the right panel of Fig. \ref{Fig2} for details). When the planet is located at the KL fixed point, orbital elements including the eccentricity, inclination and argument of pericentre remain stationary. In this special case, the Hamiltonian governing stellar spin given by equation (\ref{Eq7}) becomes
\begin{equation}\label{Eq9}
	\begin{aligned}
		{\cal H} =&  - \frac{{3{{\cal C}_1}}}{{2{{\cal C}_0}{S^*}G_k^3}}{p^2} + \frac{{12H_k^2}}{{L_p^2G_k^3}}\left( {5L_p^2 - 4G_k^2} \right)p\\
		& - \frac{{12{H_k}}}{{L_p^2G_k^3}}\sqrt {\left( {1 - {p^2}} \right)\left( {G_k^2 - H_k^2} \right)} \left( {5L_p^2 - 4G_k^2} \right)\cos \phi,
	\end{aligned}
\end{equation}
where $G_k = L_p\sqrt{1-e_k^2}$ and $H_k = G_k\cos{i_k}$ are constant.

\begin{figure*}
	\centering
	\includegraphics[width=\columnwidth]{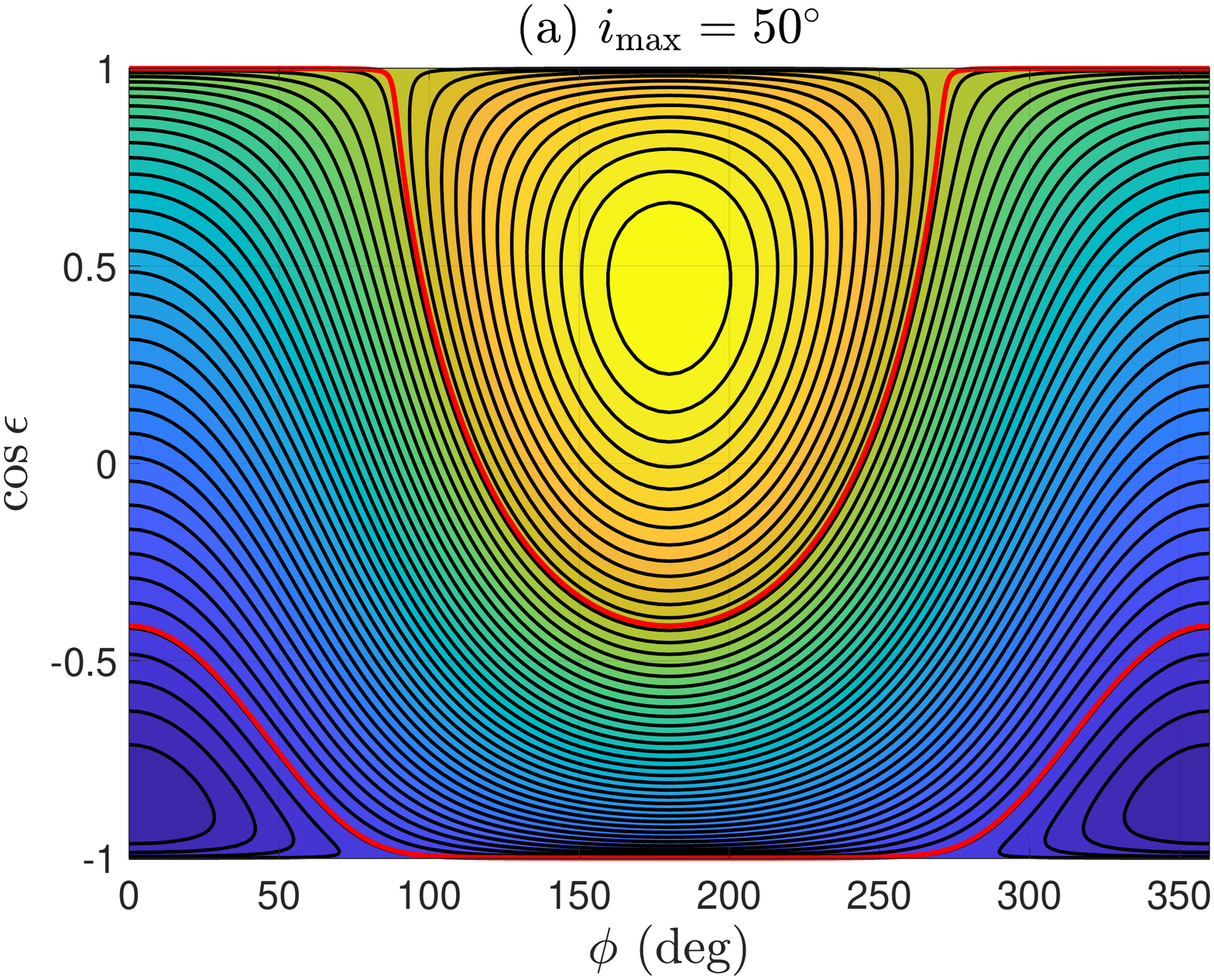}
	\includegraphics[width=\columnwidth]{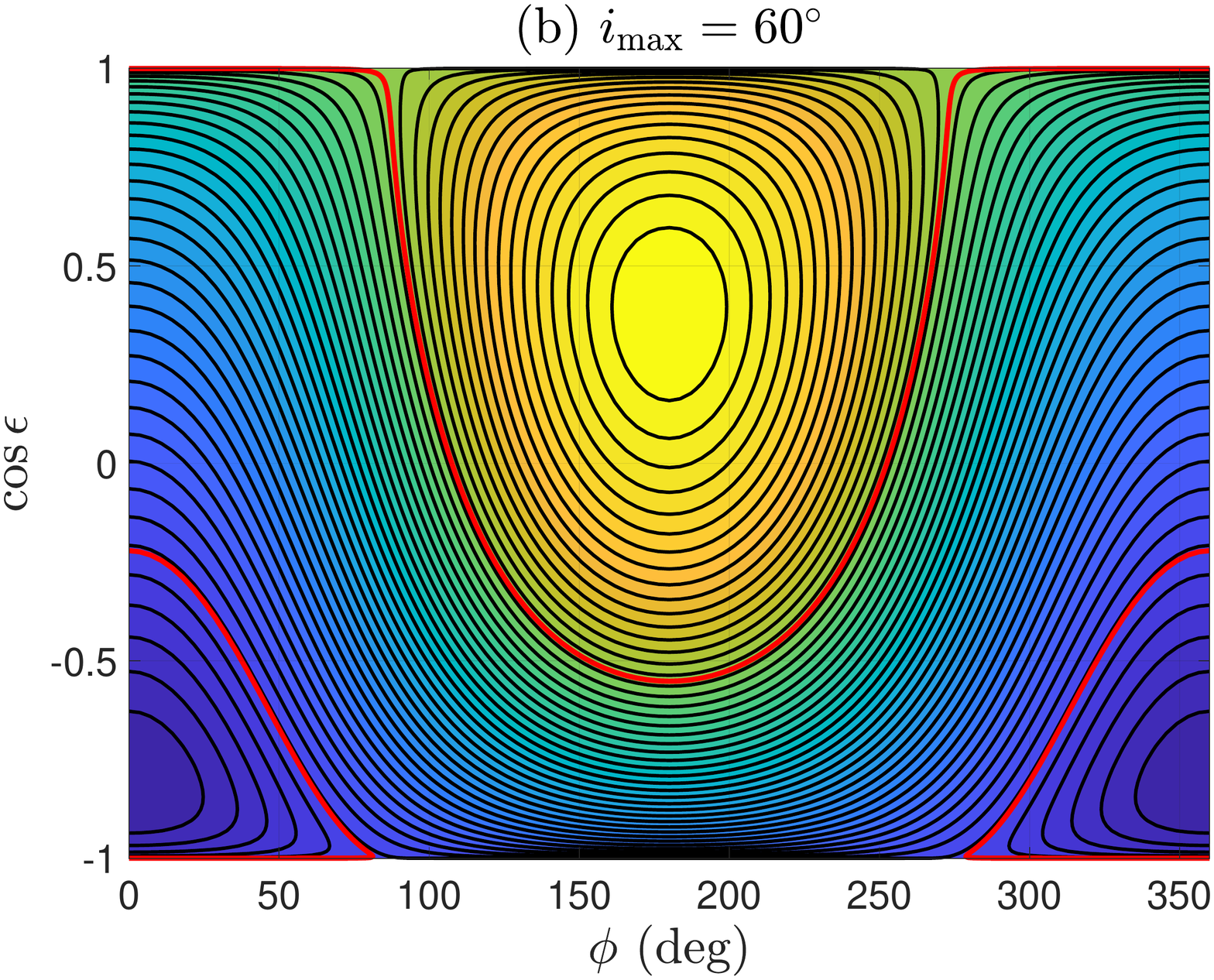}\\
	\includegraphics[width=\columnwidth]{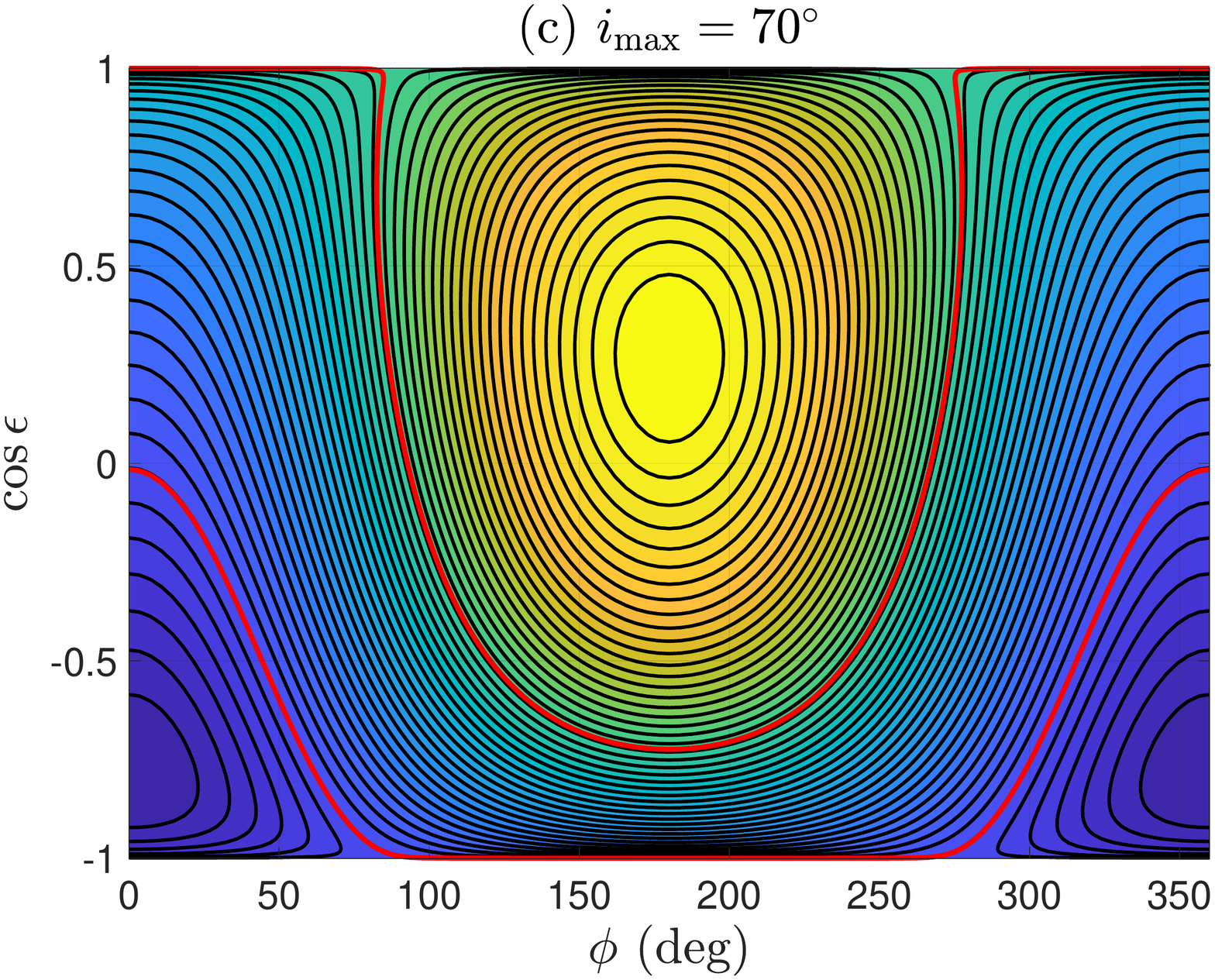}
	\includegraphics[width=\columnwidth]{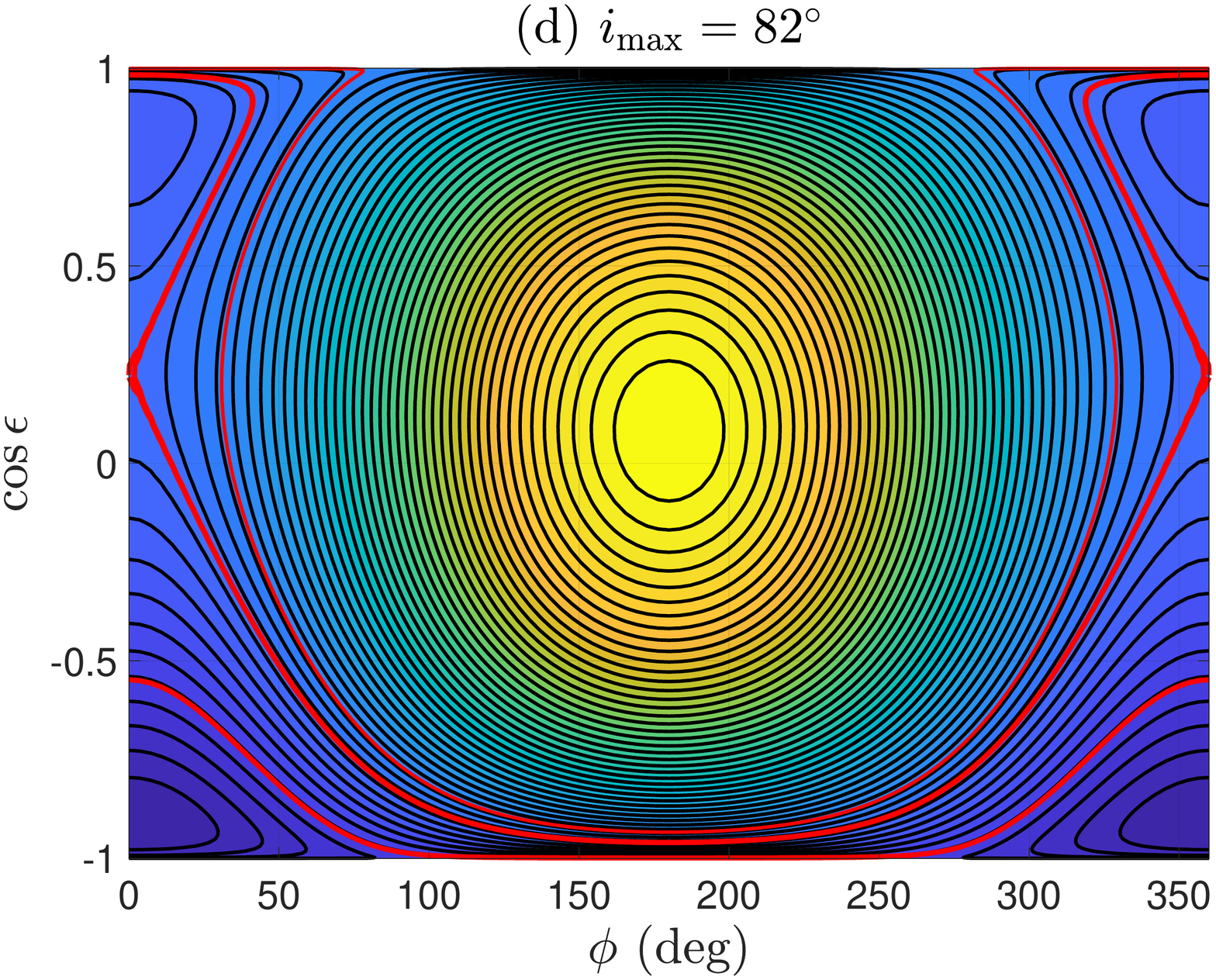}
	\caption{Phase-space structures of stellar spin dynamics when the planet is located at the KL centre. Dynamical separatrices are shown in red lines. Here, the cases of $i_{\max} = 50^{\circ}$, $i_{\max} = 60^{\circ}$, $i_{\max} = 70^{\circ}$ and $i_{\max} = 82^{\circ}$ are taken into account.}
	\label{Fig4}
\end{figure*}

In this special configuration, the Hamiltonian (\ref{Eq9}) determines a dynamical system with one degree of freedom. For such an ideal system, the global structures in the phase space can be explored by plotting level curves of Hamiltonian, as shown in Fig. \ref{Fig4}. From the phase portraits, it is observed that the structures are dependent upon the motion integral characterised by $i_{\max}$. When the maximum inclinations are $i_{\max}=50^{\circ}$, $i_{\max}=60^{\circ}$ and $i_{\max}=70^{\circ}$, the basic structures arising in the phase portraits are similar and they show that there are two libration centres: one is located at $\phi = 0^{\circ}$ with $\epsilon$ greater than $90^{\circ}$ and the other one is located at $\phi = \pi$ with $\epsilon$ smaller than $90^{\circ}$. However, it becomes different for the case of $i_{\max} = 82^{\circ}$. Besides the libration centres shown in the former cases, an additional libration centre appears at $\phi = 0^{\circ}$ with $\epsilon$ smaller than $90^{\circ}$. In all the panels shown in Fig. \ref{Fig4}, the dynamical separatrices, shown by red lines, play a role of dividing the whole phase space into regions of libration and circulation. In addition, we can further observe from the phase portraits that configurations with $\epsilon = 0^{\circ}$ or  $\epsilon = 180^{\circ}$ are marginally stable since they are located on or close to the separatrices.

Under the dynamical model governed by the Hamiltonian (\ref{Eq9}), equilibrium points are determined by the following stationary conditions:
\begin{equation*}
	\dot \phi  = \frac{{\partial {\cal H}}}{{\partial p}} = 0,\quad \dot p =  - \frac{{\partial {\cal H}}}{{\partial \phi }} = 0.
\end{equation*} 
Equilibrium points are commonly called ``Cassini's states'' \citep{peale1969generalized} but here it is for the central star rather than for planets. The second condition shows that equilibrium points are located at $\phi = 0$ or $\phi = \pi$. Stable equilibrium points correspond to libration centres. By analysing the phase structures shown in Fig. \ref{Fig4}, we can further measure the resonant width by evaluating the distance between separatrices at the libration centre. 

\begin{figure*}
	\centering
	\includegraphics[width=\columnwidth]{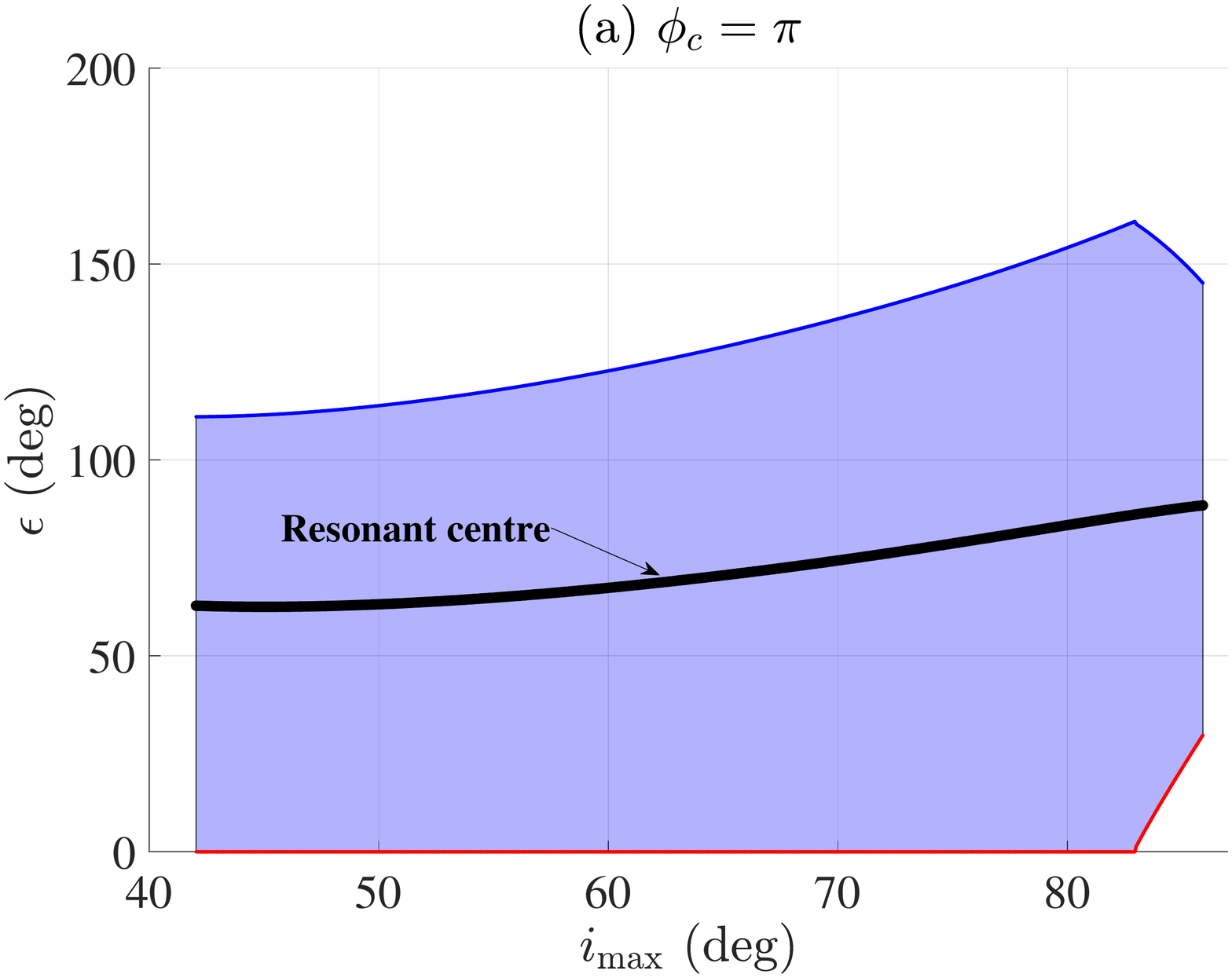}
	\includegraphics[width=\columnwidth]{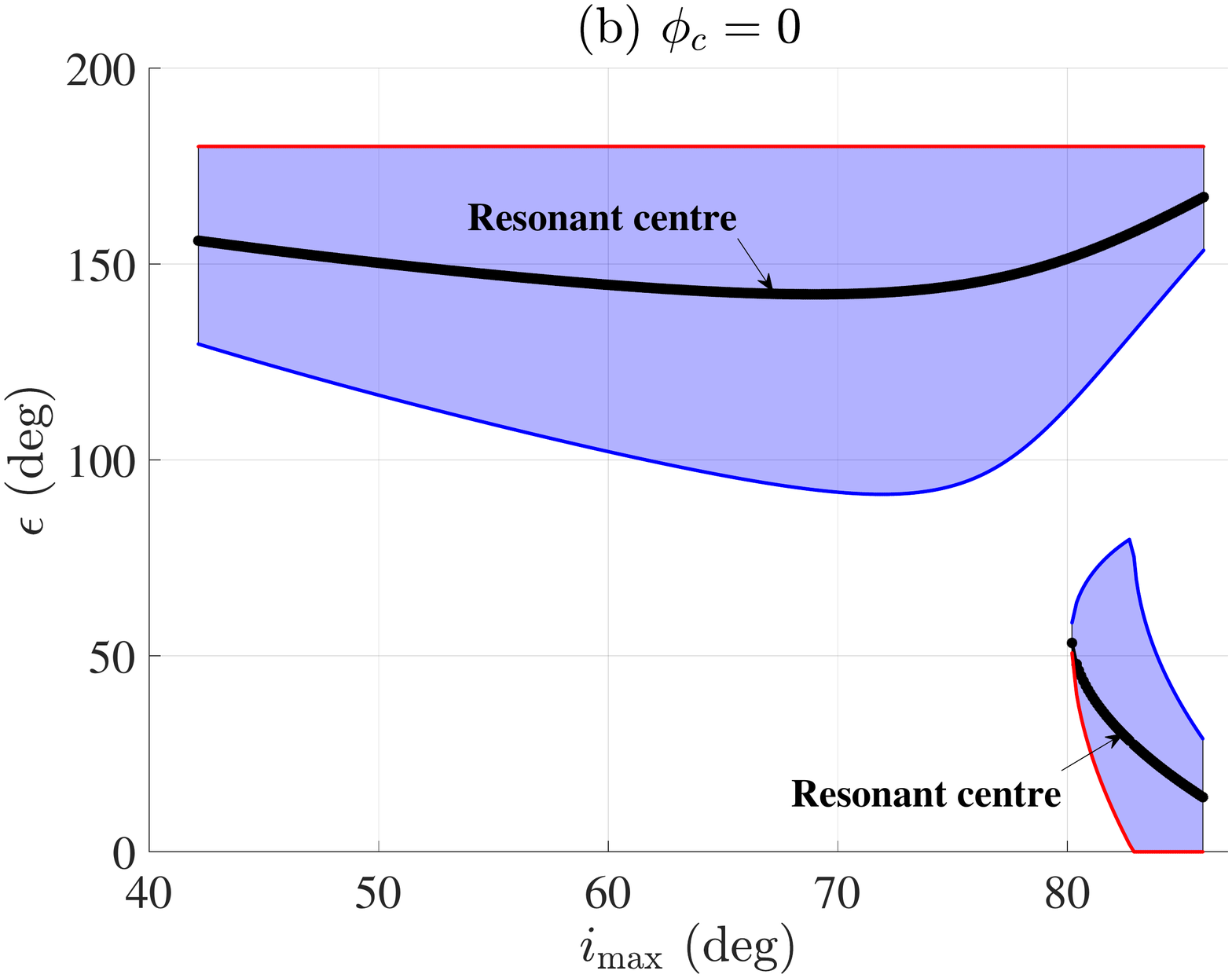}
	\caption{Resonant width as a function of $i_{\max}$ for resonant centres at $\phi_c = \pi$ (\emph{left panel}) and $\phi_c = 0$ (\emph{right panel}) when the planet is located at KL centre. The existence of KL resonance requires that $i_{\max}$ should be larger than $42.9^{\circ}$. The black lines show the distribution of resonant centres, and shaded regions stand for the resonant zones.}
	\label{Fig5}
\end{figure*}

Figure \ref{Fig5} shows the distribution of libration centres by black lines and the resonant zones by shaded areas as functions of $i_{\max}$. For libration centres at $\phi_c = \pi$, the resonant width first increases and then decreases with $i_{\max}$. For libration centres at $\phi_c = 0$, there are two branches: one is located in retrograde region of $\epsilon$ and the other one is in the prograde region of $\epsilon$. For the first branch, it exists in the entire domain of $i_{\max}$ and its resonant width first increases and then decreases with $i_{\max}$. The second branch begins to appear when the maximum inclination is greater than $\sim$$80^{\circ}$ and its resonant width first increases and then decreases with $i_{\max}$.

\section{Spin dynamics at KL librating cycles}
\label{Sect4}

\begin{figure*}
	\includegraphics[width=\columnwidth]{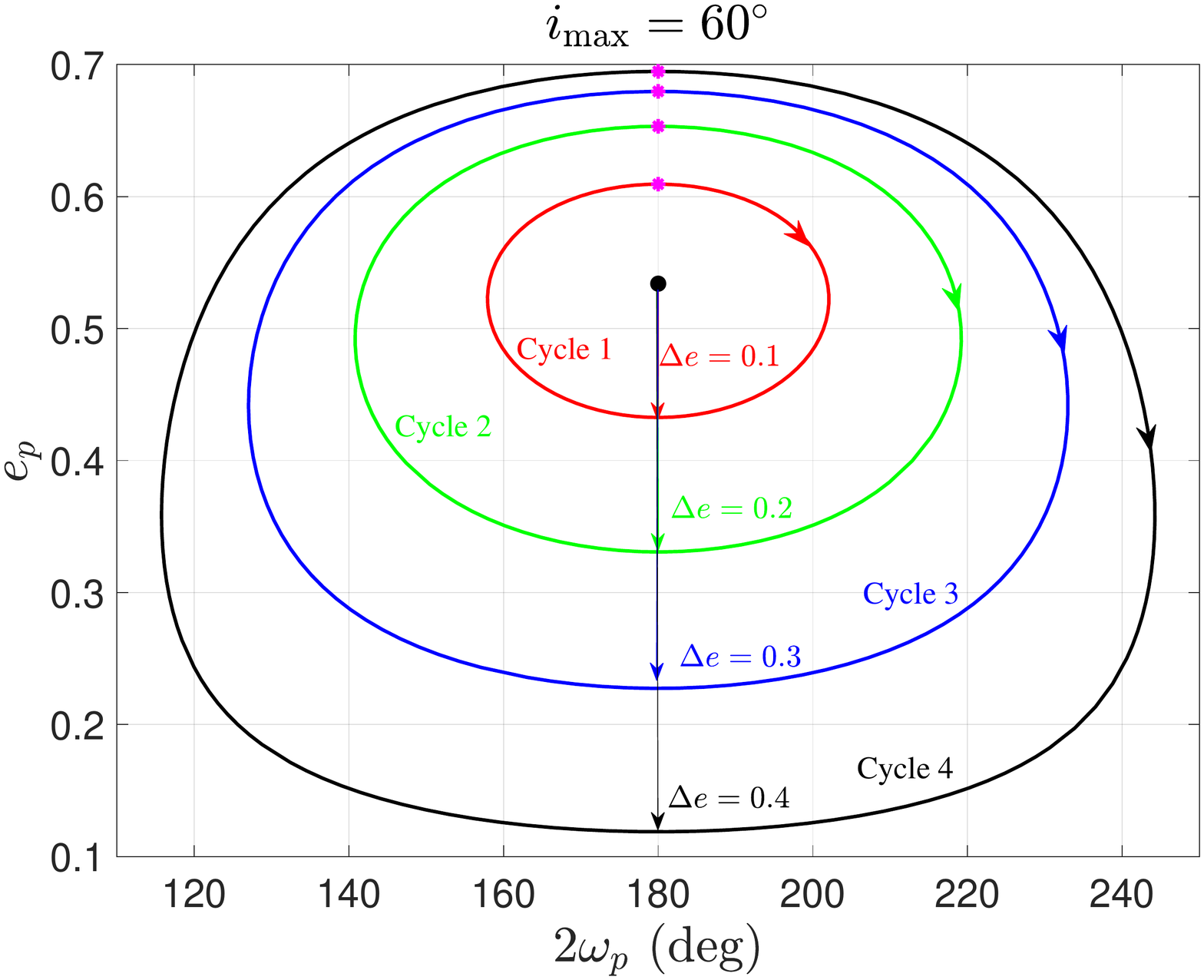}
	\includegraphics[width=\columnwidth]{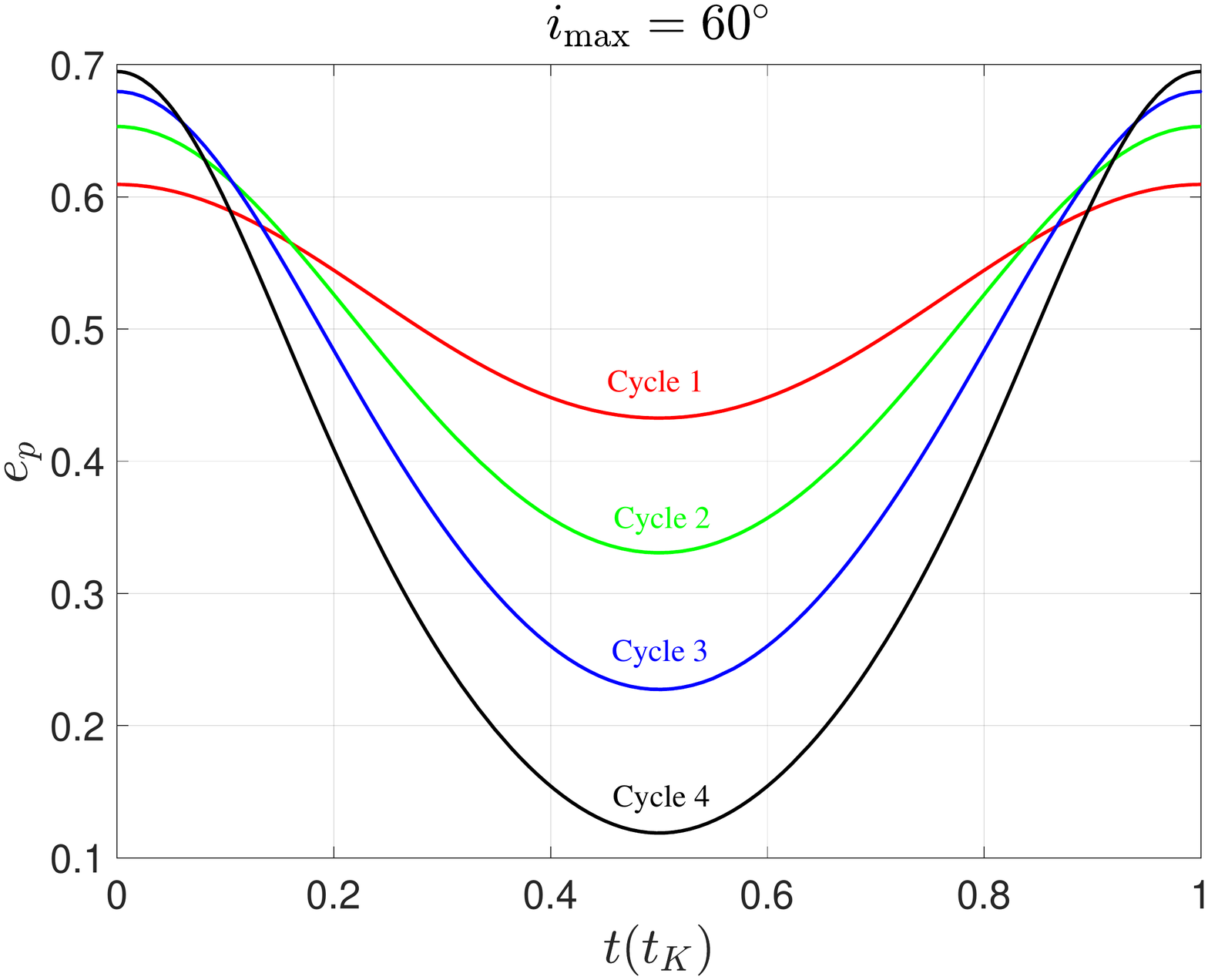}
	\caption{Representative KL cycles specified by $\Delta e = 0.1$ (red line, cycle 1), $\Delta e = 0.2$ (green line, cycle 2), $\Delta e = 0.3$ (blue line, cycle 3) and $\Delta e = 0.4$ (black line, cycle 4) with the motion integral specified by $i_{\max} = 60^{\circ}$. These cycles are periodic and they remain unchanged during the evolution of stellar spin. In the long-term spin-orbit evolution, the planet is assumed to move along these typical KL cycles. In the left panel, the starting points (i.e., the points at the initial moment $\tau = 0$) are marked by pink stars, and the black star corresponds to the location of KL centre at ($e_p = 0.534, 2\omega_p = \pi$). In the right panel, the time histories of eccentricity during one KL period are shown for these representative KL cycles (it should be noted that KL periods of these KL cycles are different).}
	\label{Fig6}
\end{figure*}

In this section, spin dynamics of the central star are studied in configurations where planets are moving on KL librating cycles. 

It should be mentioned that the dynamics of stellar spin in configurations with planets located outside KL resonance specified by extremely high $i_{\max}$ have been investigated by \citet{storch2014chaotic}, \citet{storch2015chaotic} and \citet{storch2017dynamics}. The KL cycles they adopted are fixed and close to the KL separatrix. Thus, it is not clear about the influence of different-size KL cycles upon stellar spin dynamics. In addition, for the non-adiabatic case, \citet{storch2017dynamics} consider the dynamical structure caused by the $N=0$ Hamiltonian. However, the contribution of higher $N$ Hamiltonian to the dynamical structures is not clear. In this work, we attempt to address these two points.

In practice, we assume that the orbits of planet follow along the KL cycles shown in Fig. \ref{Fig6}. The motion integral is specified by $i_{\max} = 60^{\circ}$ and the KL centre is located at ($e_p = 0.534, 2\omega_p = \pi$). These cycles are characterised by $\Delta e = 0.1$ (cycle 1), $\Delta e = 0.2$ (cycle 2), $\Delta e = 0.3$ (cycle 3) and $\Delta e = 0.4$ (cycle 4). Their eccentricity variations during one KL period are presented in the right panel of Fig. \ref{Fig6}. It should be emphasised that these KL cycles are exactly periodic and they are not influenced by stellar rotation. In the following, we intend to study the spin dynamics of the central star under the configurations where planets move on these typical KL librating cycles.

\subsection{Approximate Hamiltonian}
\label{Sect4-1}

\begin{figure*}
	\includegraphics[width=\columnwidth]{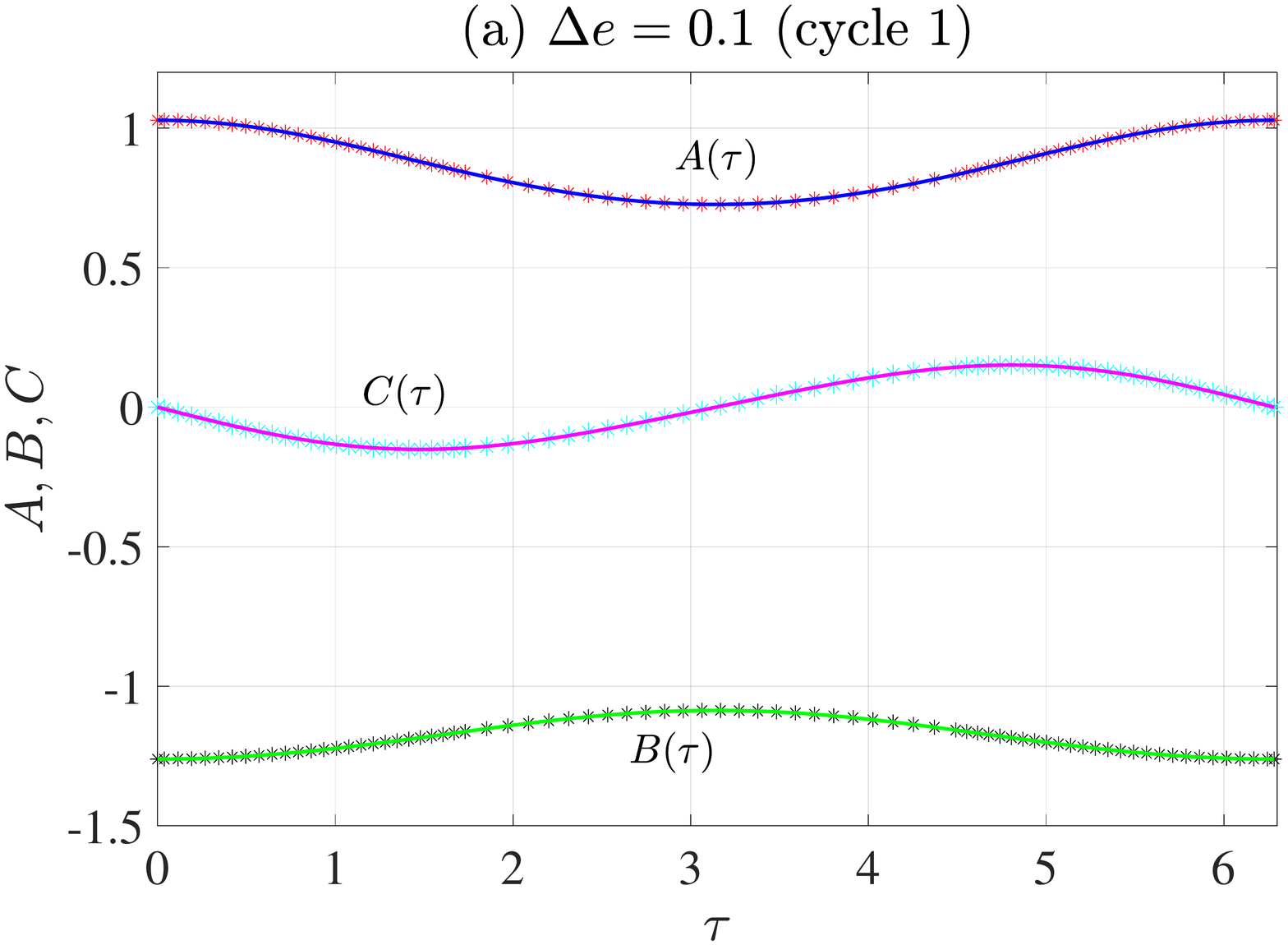}
	\includegraphics[width=\columnwidth]{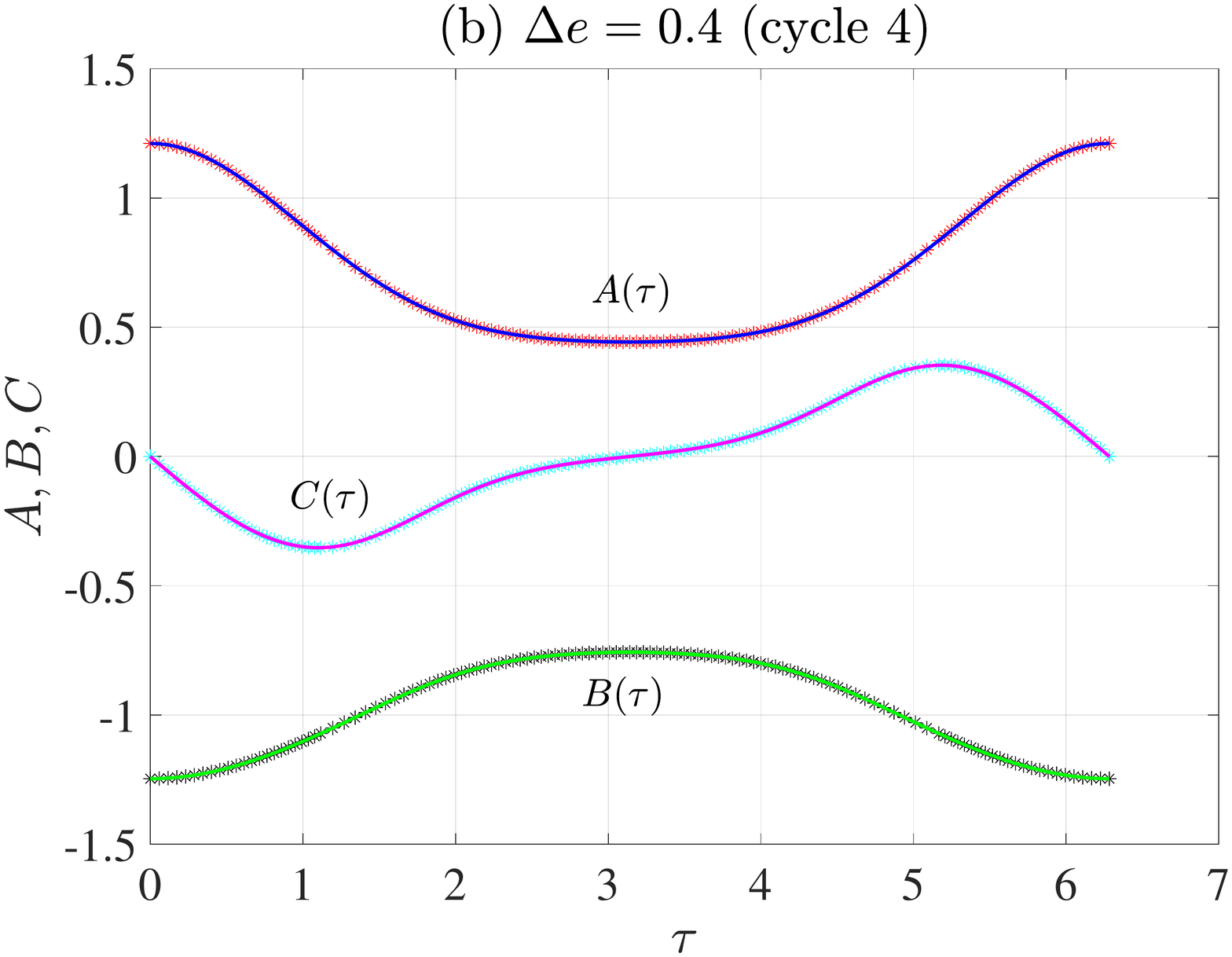}
	\caption{Accurate and approximate curves of $A(\tau)$, $B(\tau)$ and $C(\tau)$ during one KL period for cycle 1 (\emph{left panel}) and cycle 4 (\emph{right panel}). The approximate curves are produced from the Fourier fitting up to order $N = 6$.}
	\label{Fig7}
\end{figure*}

To simplify the dynamical model, let us denote the coefficient arising in the Hamiltonian (\ref{Eq7}) by $\alpha(t)$,
\begin{equation*}
	\alpha \left( t \right) = \frac{{3{{\cal C}_1}}}{{{{\cal C}_0}{S^*}G_p^3}}.
\end{equation*}
Similar to \citet{storch2015chaotic}, a scaled time variable $\tau$ is introduced by
\begin{equation}\label{Eq10}
	\tau (t) = \frac{{{n_e}}}{{\bar \alpha }}\int\limits_0^t {\alpha (t'){\rm d}t'},
\end{equation}
where the averaged value of $\alpha$ is calculated by
\begin{equation*}
	\bar \alpha  = \frac{{{n_e}}}{{2\pi }}\int\limits_0^{2\pi /{n_e}} {\alpha (t)} {\rm d}t.
\end{equation*}
In equation (\ref{Eq10}), $n_e$ is the angular frequency of KL cycle. In $\tau$ space, KL cycles shown in Fig. \ref{Fig6} have periods of $2\pi$ and the Hamiltonian (\ref{Eq7}) becomes
\begin{equation}\label{Eq11}
	\begin{aligned}
		\begin{aligned}
			{{\cal H}} =& \frac{{\bar \alpha }}{{{n_e}}}\left\{ { - \frac{1}{2}{p^2} + \frac{1}{\alpha }\frac{{6H_p^2}}{{L_p^2G_p^3}}\left[ {5L_p^2 - 3G_p^2 - 5\left( {L_p^2 - G_p^2} \right)} \right.} \right.\\
			&\times \left. {\cos 2{g_p}} \right]p + \frac{1}{\alpha }\frac{{6{H_p}}}{{L_p^2G_p^3}}\sqrt {G_p^2 - H_p^2} \sqrt {1 - {p^2}} \\
			&\times \left[ {\left( {5\left( {L_p^2 - G_p^2} \right)\cos 2{g_p} - \left( {5L_p^2 - 3G_p^2} \right)} \right)\cos \phi } \right.\\
			&\left. {\left. { + 5\left( {L_p^2 - G_p^2} \right)\sin 2{g_p}\sin \phi } \right]} \right\}.
		\end{aligned}
	\end{aligned}
\end{equation}
To simplify equation (\ref{Eq11}), we denote
\begin{equation}\label{Eq12}
	\begin{aligned}
		A(\tau ) =& \frac{1}{{\alpha (\tau )}}\frac{{6H_p^2}}{{L_p^2G_p^3}}\left[ {5L_p^2 - 3G_p^2 - 5\left( {L_p^2 - G_p^2} \right)\cos 2{g_p}} \right],\\
		B(\tau ) =& \frac{1}{{\alpha (\tau )}}\frac{{6{H_p}}}{{L_p^2G_p^3}}\sqrt {G_p^2 - H_p^2} \left[ {5\left( {L_p^2 - G_p^2} \right)\cos 2{g_p}} \right.\\
		&\left. { - \left( {5L_p^2 - 3G_p^2} \right)} \right],\\
		C(\tau ) =& \frac{1}{{\alpha (\tau )}}\frac{{30{H_p}}}{{L_p^2G_p^3}}\sqrt {G_p^2 - H_p^2} \left( {L_p^2 - G_p^2} \right)\sin 2{g_p}.
	\end{aligned}
\end{equation}
By replacing equation (\ref{Eq12}) in the Hamiltonian (\ref{Eq11}), we can obtain
\begin{equation}\label{Eq13}
	\begin{aligned}
		{{\cal H}} =& \frac{{\bar \alpha }}{{{n_e}}}\left[ { - \frac{1}{2}{p^2} + A(\tau )p} \right.\\
		&\left. { + \sqrt {1 - {p^2}} \left[ {B(\tau )\cos \phi  + C(\tau )\sin \phi } \right]} \right].
	\end{aligned}
\end{equation}
The coefficients $A(\tau)$, $B(\tau)$ and $C(\tau)$ are periodic and their periods are $2\pi$ in $\tau$ space, equal to the period of KL cycle. Similar to \citet{storch2015chaotic}, $A(\tau)$, $B(\tau)$ and $C(\tau)$ can be decomposed as Fourier series (truncated at order $N$) as follows:
\begin{equation}\label{Eq14}
	\begin{aligned}
		A(\tau ) =& \sum\limits_{n = 0}^N {{A_n}\cos \left( {n\tau } \right)}, \\
		B(\tau ) =& \sum\limits_{n = 0}^N {{B_n}\cos \left( {n\tau } \right)}, \\
		C(\tau ) =& \sum\limits_{n = 1}^N {{C_n}\sin \left( {n\tau } \right)}, 
	\end{aligned}
\end{equation}
where $A_n$, $B_n$ and $C_n$ are coefficients depending on the `shape' of KL cycle. Figure \ref{Fig7} shows the curves of $A(\tau)$, $B(\tau)$ and $C(\tau)$ produced from the original expression (\ref{Eq12})  and the Fourier decomposition (\ref{Eq14}) up to order 6 for KL cycle 1 in the left panel and KL cycle 4 in the right panel. One can see that the Fourier decomposition can reproduce the curves of $A$, $B$ and $C$ in an adequate accuracy.

By replacing the Fourier decomposition (\ref{Eq14}) in equation (\ref{Eq13}), the Hamiltonian becomes
\begin{equation}\label{Eq15}
	\begin{aligned}
		{\cal H} =& \frac{{\bar \alpha }}{{{n_e}}}\left( { - \frac{1}{2}{p^2} + {A_0}p + {B_0}\sqrt {1 - {p^2}} \cos \phi } \right) + \frac{{\bar \alpha }}{{{n_e}}}\\
		&  \times \left\{ {p\sum\limits_{n \ge 1} {{A_n}\cos \left( {n\tau } \right){\rm{ + }}\frac{1}{2}\sqrt {1 - {p^2}} } } \right.{\kern 1pt} \sum\limits_{n \ge 1} {\left[ {\left( {{B_n} + {C_n}} \right)} \right.} {\kern 1pt} \\
		&\left. {\left. { \times \cos \left( {\phi  - n\tau } \right) + \left( {{B_n} - {C_n}} \right)\cos \left( {\phi  + n\tau } \right)} \right]} \right\}.
	\end{aligned}
\end{equation}
The Hamiltonian (\ref{Eq15}) is dependent on $\tau$, showing that it is non-autonomous. An equivalent Hamiltonian can be found in \citet{storch2015chaotic} under a different notation system (see equation (44) in their study).

In order to study high-order as well as secondary spin-orbit resonances, the Hamiltonian system is augmented into a 2-DOF one by introducing an action variable $T$ which is conjugated to $\tau$ as follows (here we still use ${\cal H}$ to denote the augmented Hamiltonian):
\begin{equation}\label{Eq16}
	\begin{aligned}
		{\cal H} =&\; T + \frac{{\bar \alpha }}{{{n_e}}}\left( { - \frac{1}{2}{p^2} + {A_0}p + {B_0}\sqrt {1 - {p^2}} \cos \phi } \right)\\
		& + \frac{{\bar \alpha }}{{{n_e}}}\left\{ {p\sum\limits_{n \ge 1} {{A_n}\cos \left( {n\tau } \right)}  + \frac{1}{2}\sqrt {1 - {p^2}}  \times } \right.\\
		&\left. {\sum\limits_{n \ge 1} {\left[ {\left( {{B_n} + {C_n}} \right)\cos \left( {\phi  - n\tau } \right) + \left( {{B_n} - {C_n}} \right)\cos \left( {\phi  + n\tau } \right)} \right]} } \right\}.
	\end{aligned}
\end{equation}
Hamiltonian canonical relations lead to the equations of motion, expressed by
\begin{equation}\label{Eq17}
	\begin{aligned}
		{\phi}' =& \frac{{\partial {\cal H}}}{{\partial p}},\quad p' =  - \frac{{\partial {\cal H}}}{{\partial \phi }},\\
		{\tau}' =&  \frac{{\partial {\cal H}}}{{\partial T}} = 1.0,\quad T' =  - \frac{{\partial {\cal H}}}{{\partial \tau }},
	\end{aligned}
\end{equation}
where $x'$ denotes the derivative of $x$ with respect to $\tau$. As for the equations of motion, equation (\ref{Eq8}) is given in time space and (\ref{Eq17}) is given in $\tau$ space. However, both versions of equations of motion are equivalent.

\begin{figure*}
	\centering
	\includegraphics[width=\columnwidth]{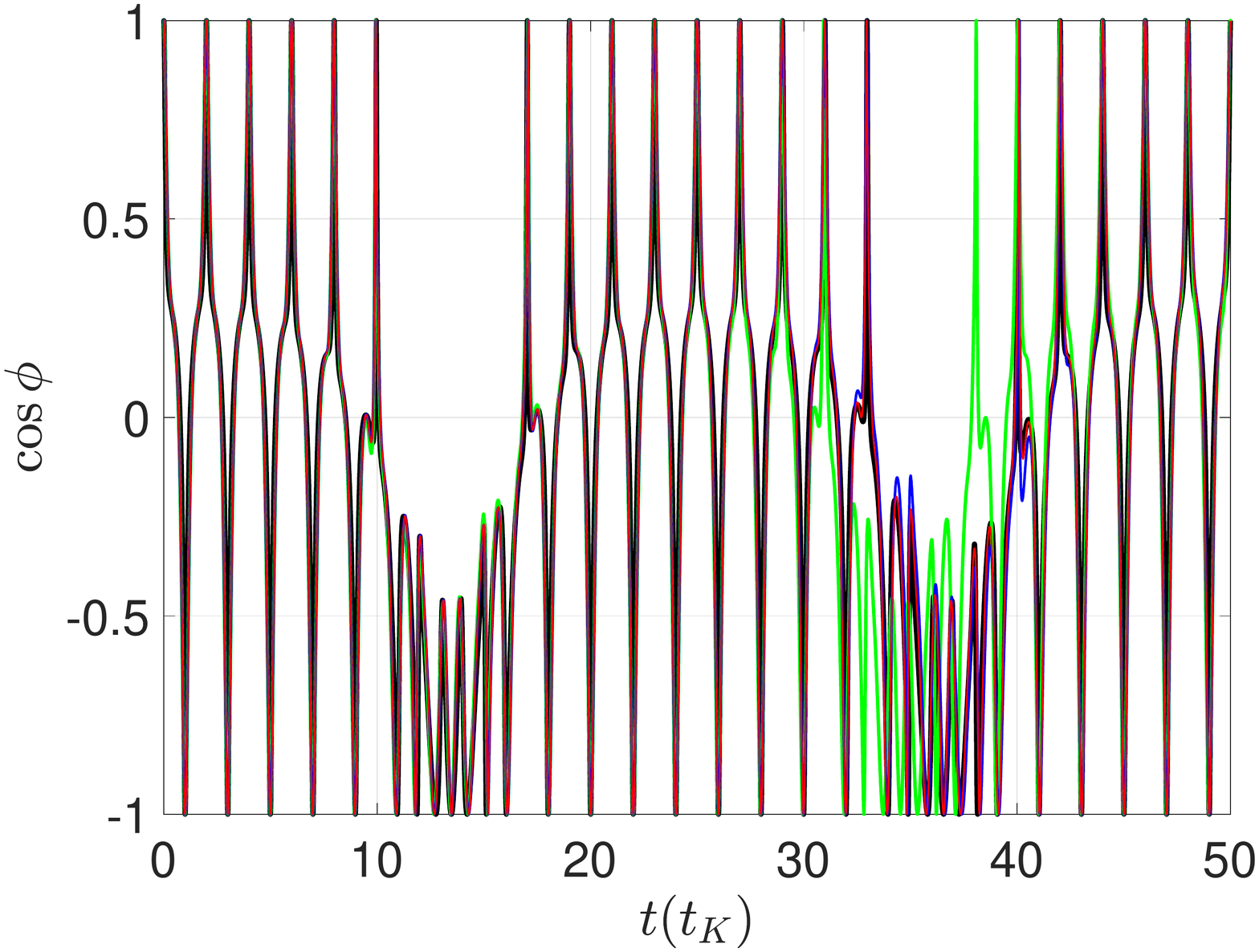}
	\includegraphics[width=\columnwidth]{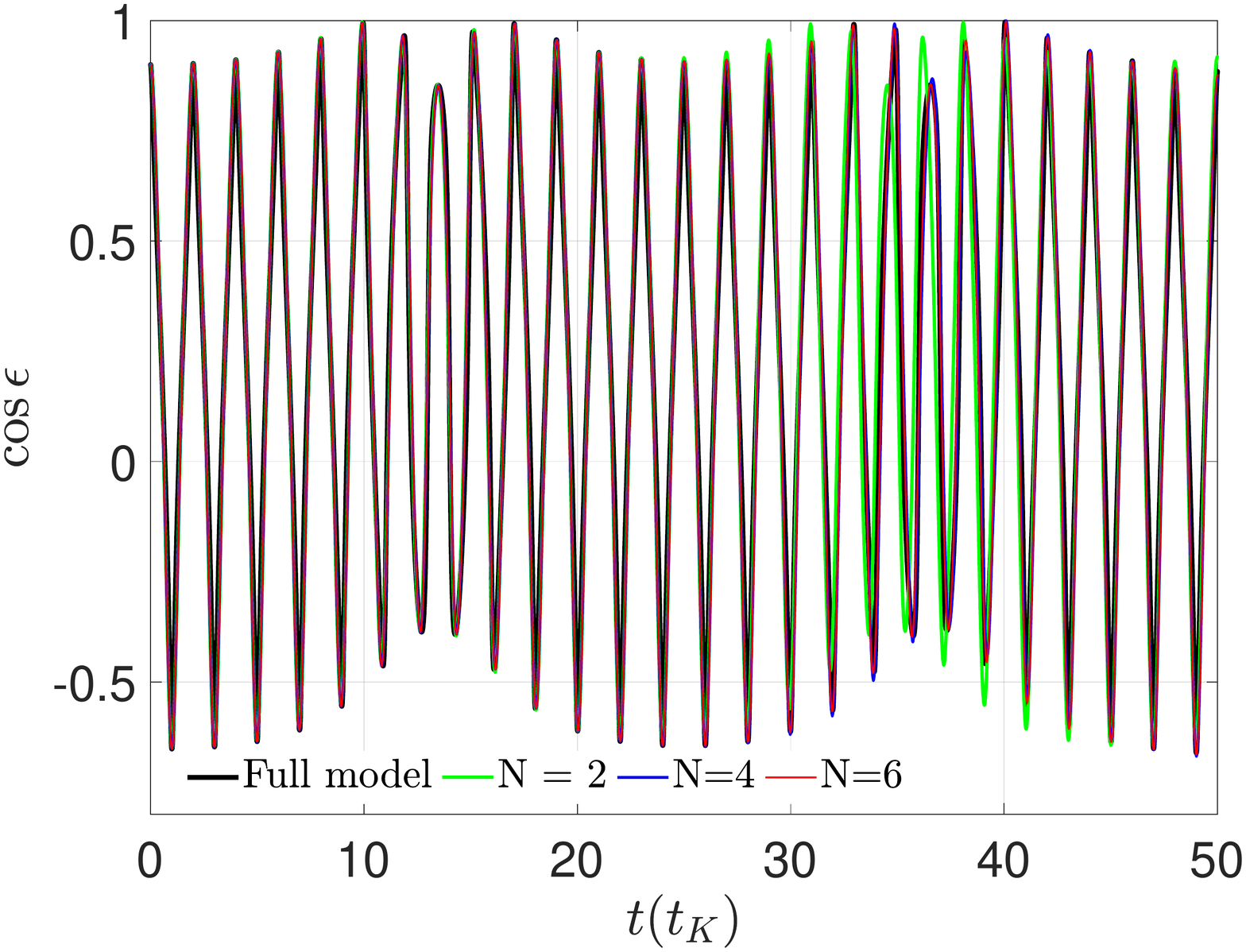}
	\caption{Evolution of stellar spin-orbit states propagated under the full spin model (black line) and under the approximated model up to order 2 (green line), 4 (blue line), 6 (red line).}
	\label{Fig8}
\end{figure*}

For convenience, we referred to equation (\ref{Eq8}) as the full model and equation (\ref{Eq17}) as the approximated model. In Fig. \ref{Fig8}, the equations of motion under the full model and approximated modes up to different orders are numerically integrated over 50 KL periods by starting from the same initial condition. As expected, the approximated model up to a higher order can approximate the full model better. Especially, the 6th-order model works very well.

In the following simulations, we will discuss secular dynamics of stellar spin under the 6th-order approximate model, governed by the Hamiltonian (\ref{Eq16}). In particular, the numerical technique of Poincar\'e section is taken in Section \ref{Sect4-2} to explore the global dynamical structures in the phase space. Then, perturbative treatments are adopted in Sections \ref{Sect4-3} and \ref{Sect4-4} to analyse the dynamical mechanism causing complex structures of spin.

\subsection{Poincar\'e sections}
\label{Sect4-2}

\begin{figure*}
	\centering
	\includegraphics[width=\columnwidth]{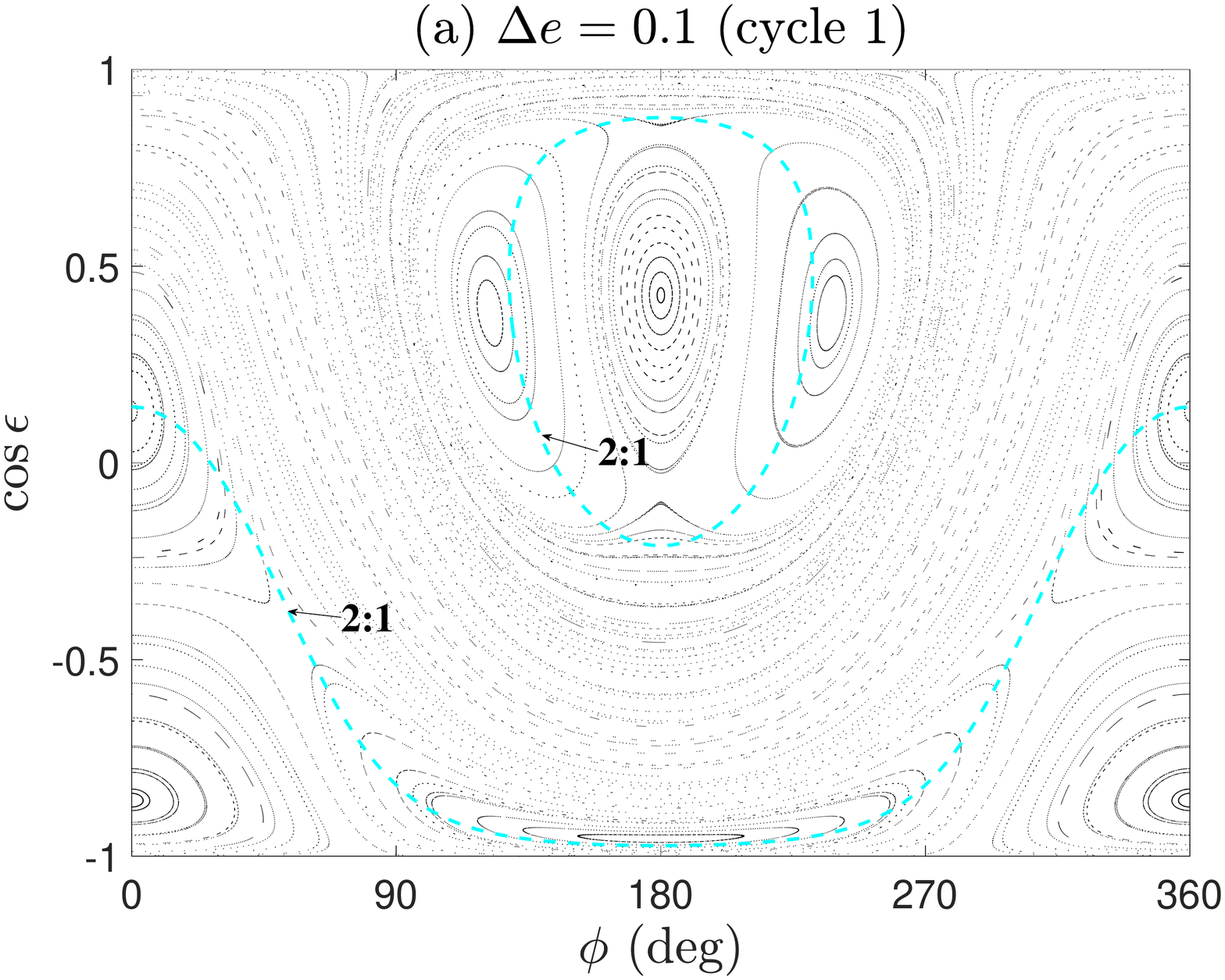}
	\includegraphics[width=\columnwidth]{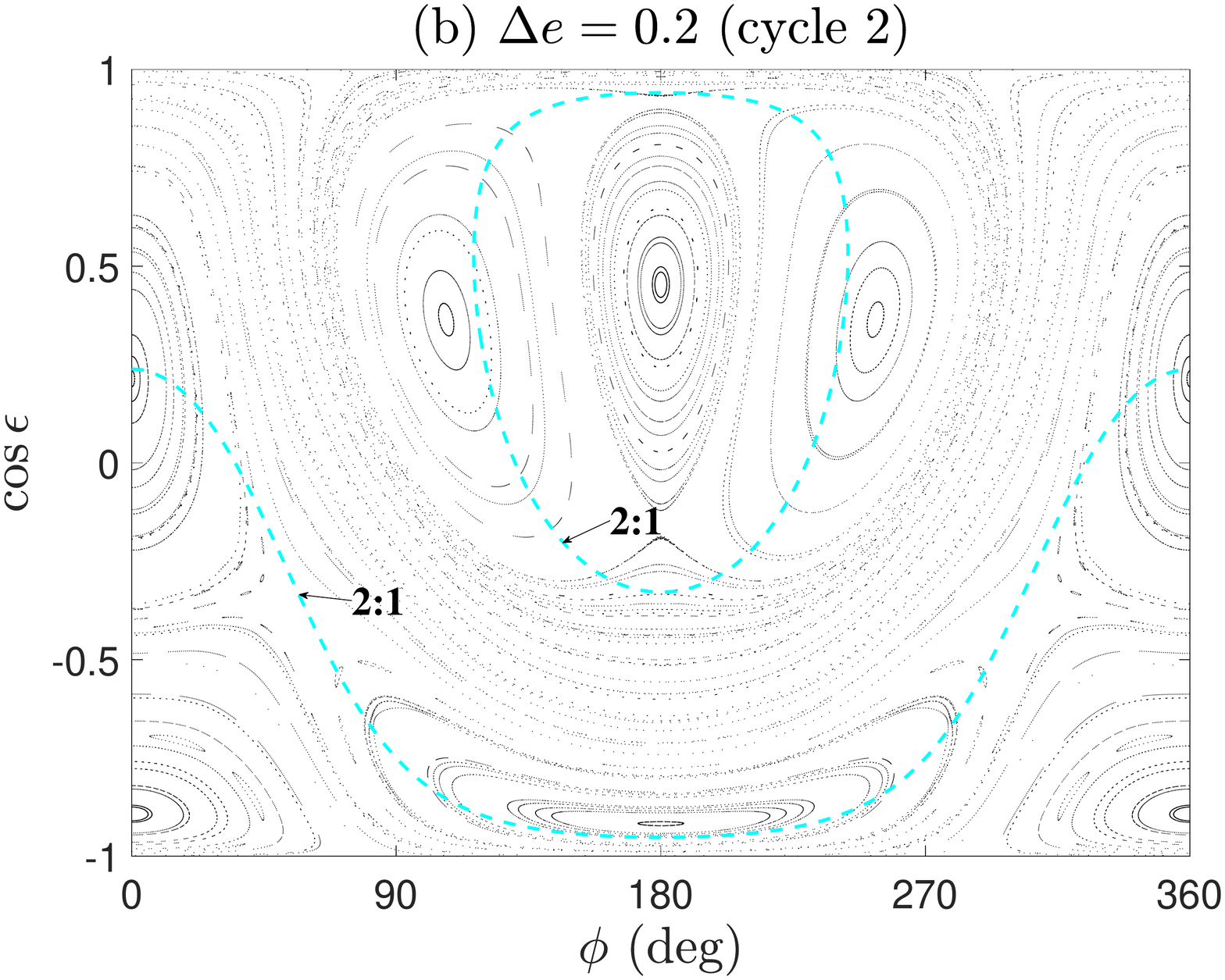}\\
	\includegraphics[width=\columnwidth]{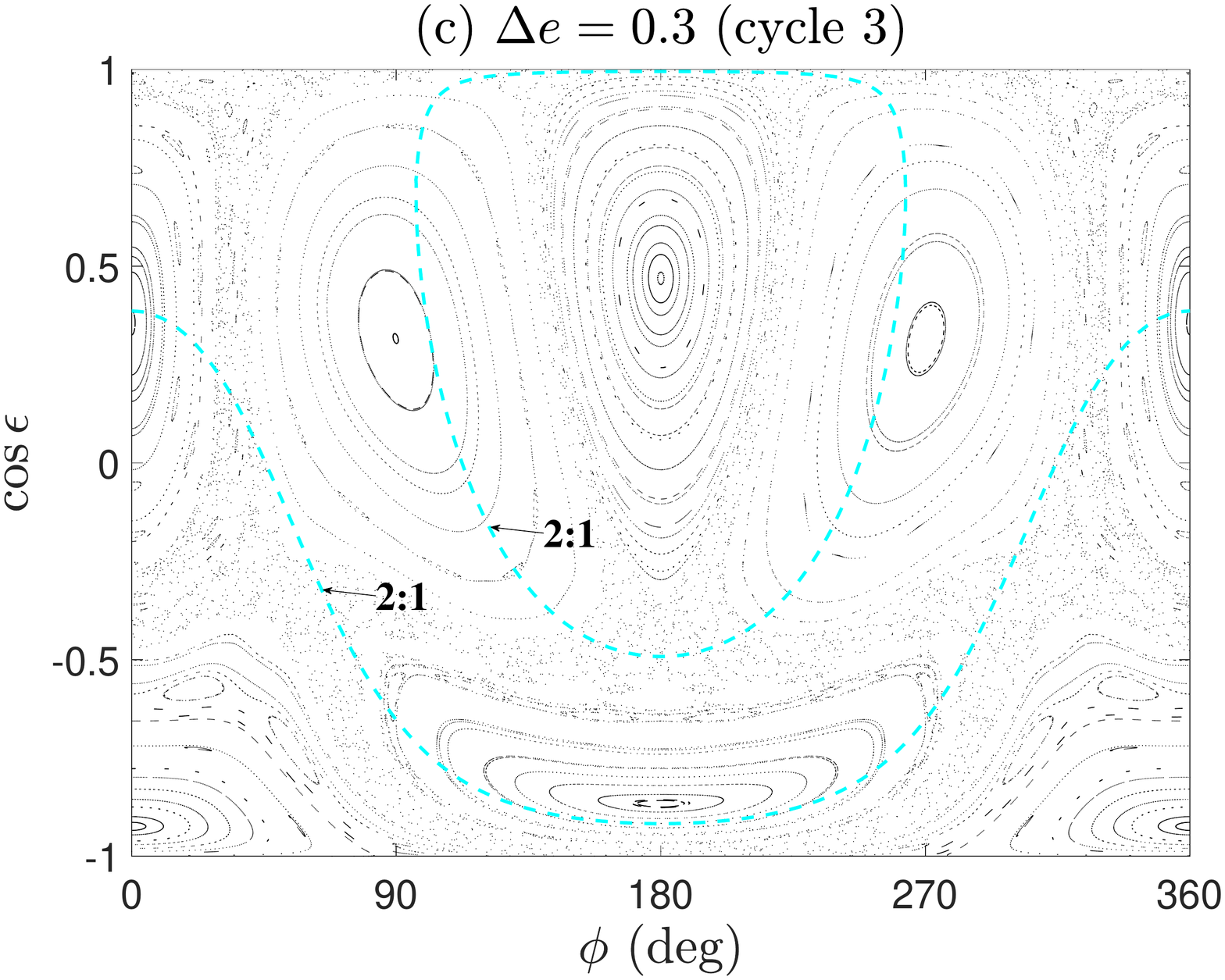}
	\includegraphics[width=\columnwidth]{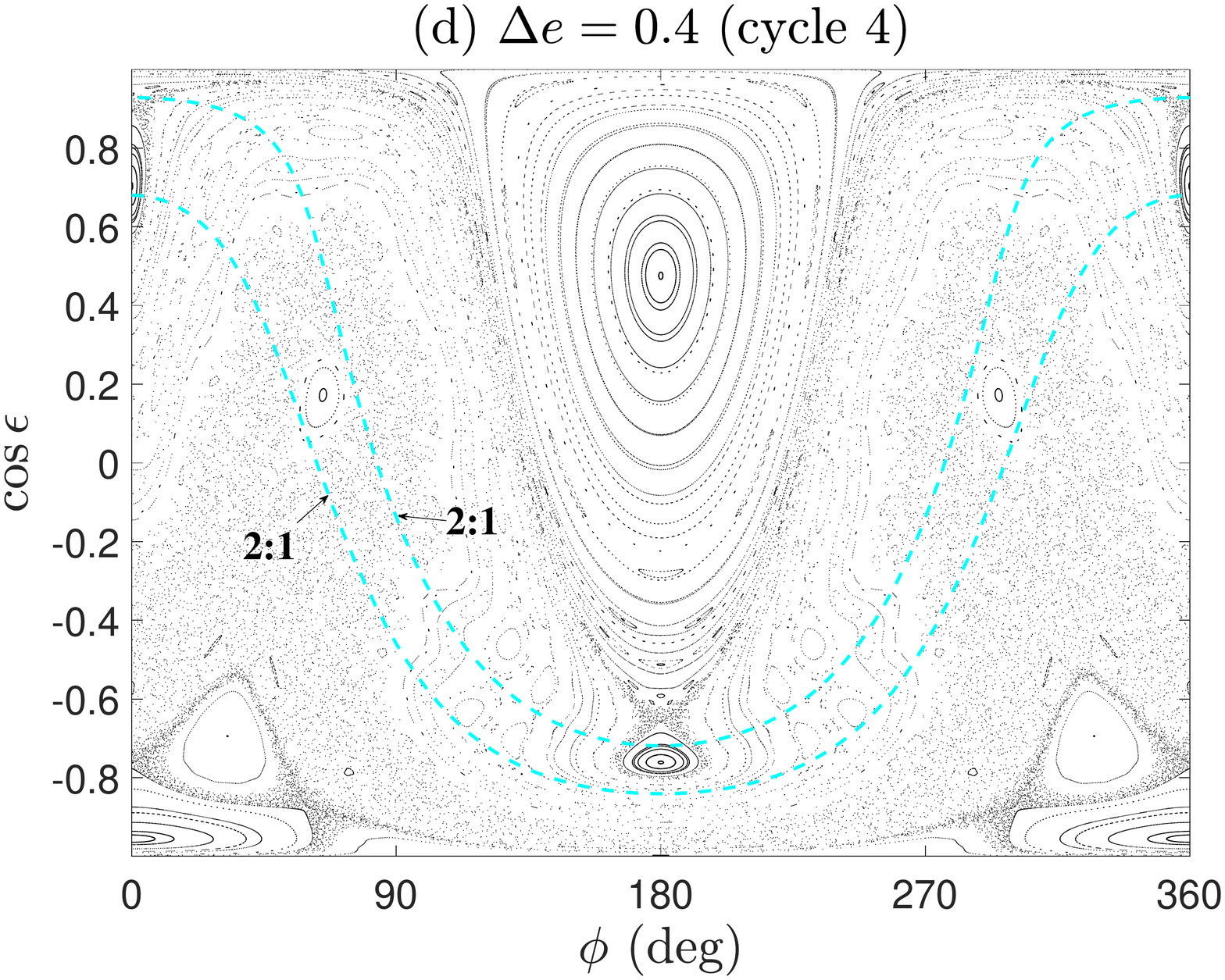}
	\caption{Poincar\'e sections defined by $g_p = \pi/2$ and $\dot g_p > 0$ when the planet is moving on KL cycle 1 (\emph{upper-left panel}), cycle 2 (\emph{upper-right panel}), cycle 3 (\emph{bottom-left panel}) and cycle 4 (\emph{bottom-right panel}). The cyan dashed lines stand for the nominal locations of 2:1 spin-orbit (high-order and/or secondary) resonances (see the descriptions in Sect. \ref{Sect4-3}).}
	\label{Fig9}
\end{figure*}

The Hamiltonian (\ref{Eq16}) determines a two-degree-of-freedom dynamical model. For such a 2-DOF model, Poincar\'e section is a powerful tool to investigate global structures in phase space. To this end, the following Poincar\'e section is defined:
\begin{equation}\label{Eq18}
	g_p = \pi/2,\quad {\dot g}_p > 0.
\end{equation}
We can see that, at the points of section, the eccentricity reaches its maximum during the long-term evolution (see Fig. \ref{Fig6}). 

In $\tau$ space, it is much easier to produce Poincar\'e sections by recording spin states at the moments when
\begin{equation*}
	\mod(\tau,2\pi) = 0.
\end{equation*}

Figure \ref{Fig9} shows Poincar\'e sections for the configurations in which planets are moving on the KL cycles shown in Fig. \ref{Fig6}. In Poincar\'e sections, those continuous lines stand for regular motions and those scattered points are chaotic trajectories. In particular, those continuous lines inside an island stand for quasi-periodic trajectories,  and the centre of an island (resonant centre) corresponds to a periodic orbit. Thanks to Poincr\'e sections, we can understand global structures of stellar spin in configurations with planets moving on different KL cycles. 

In the case of $\Delta e = 0.1$ (cycle 1), the whole phase space is mainly filled by regular orbits. The chaotic layers existing between different islands of resonance are too narrow to observe. There are two islands centred at $\phi=0$ and there is a primary island centred at $\phi = \pi$. Inside the primary resonance, islands of secondary resonances appears and there are three sub-islands inside the primary island. The Poincar\'e sections in the cases of $\Delta e = 0.2$ (cycle 2) and $\Delta e = 0.3$ (cycle 3) hold similar structures. However, chaotic layer becomes wider with increasing $\Delta e$. This is because the perturbation becomes stronger when the KL cycle has a larger amplitude. In the case of $\Delta e = 0.3$ (cycle 3), multiple periodic islands corresponding to high-order spin-orbit resonances can be found.

It is different for the case of $\Delta e = 0.4$ (cycle 4). In this case, the phase space is filled mainly by chaotic sea (see the last panel of Fig. \ref{Fig9}).  However, the primary island of resonance centred at $\phi = \pi$ still exists. In the chaotic sea, periodic islands corresponding to high-order spin-orbit resonances can be found. 

About the complex dynamics of stellar rotation, we may ask: what is the dynamical mechanism governing the basic structures arising in the Poincar\'e sections? In the following, we take Hamiltonian perturbation theory developed by \citet{henrard1986perturbation} and \citet{henrard1990semi} to deal with this problem and attempt to provide some preliminary understanding. 

\subsection{Dynamics under the unperturbed Hamiltonian model}
\label{Sect4-3}

\begin{figure*}
	\includegraphics[width=\columnwidth]{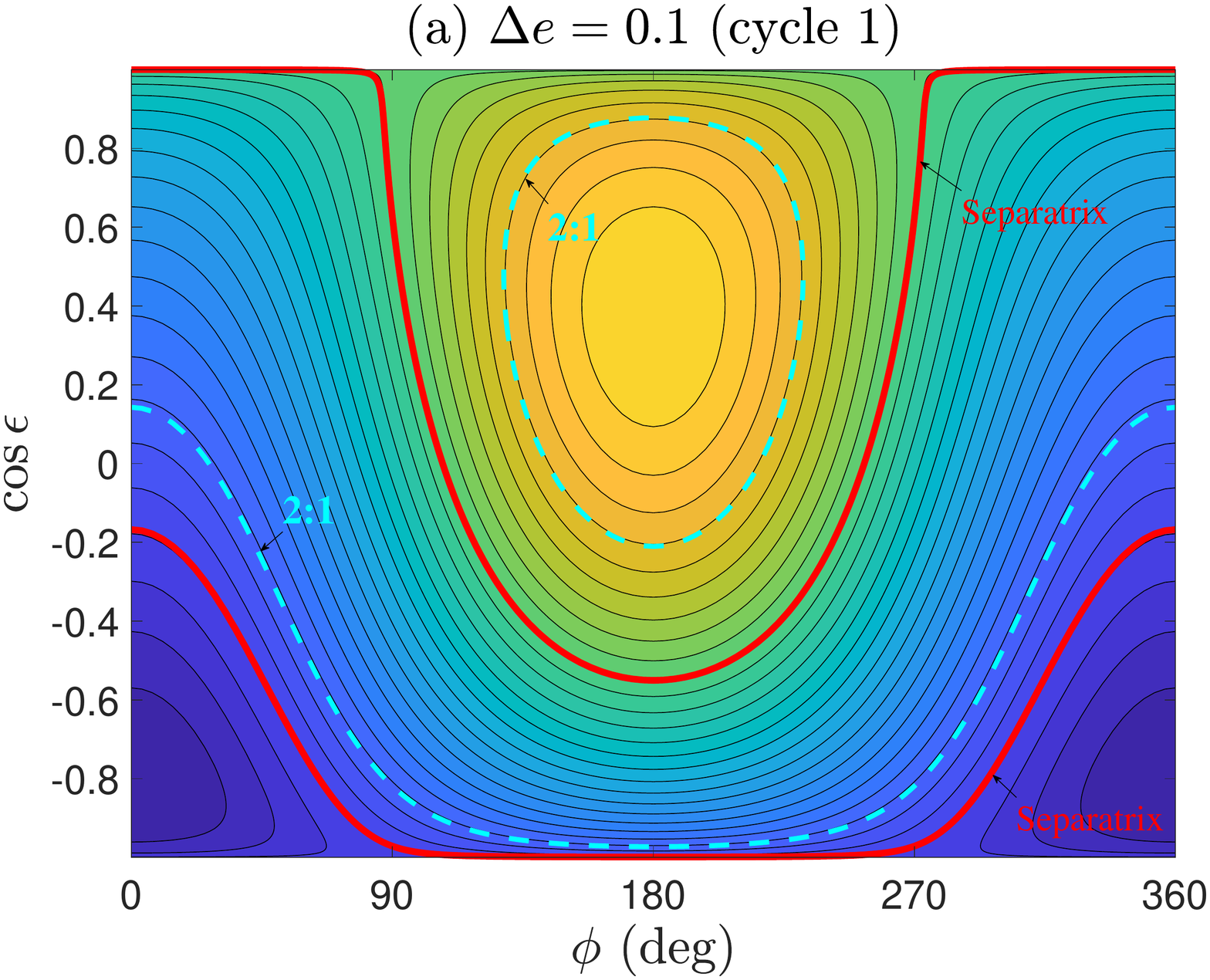}
	\includegraphics[width=\columnwidth]{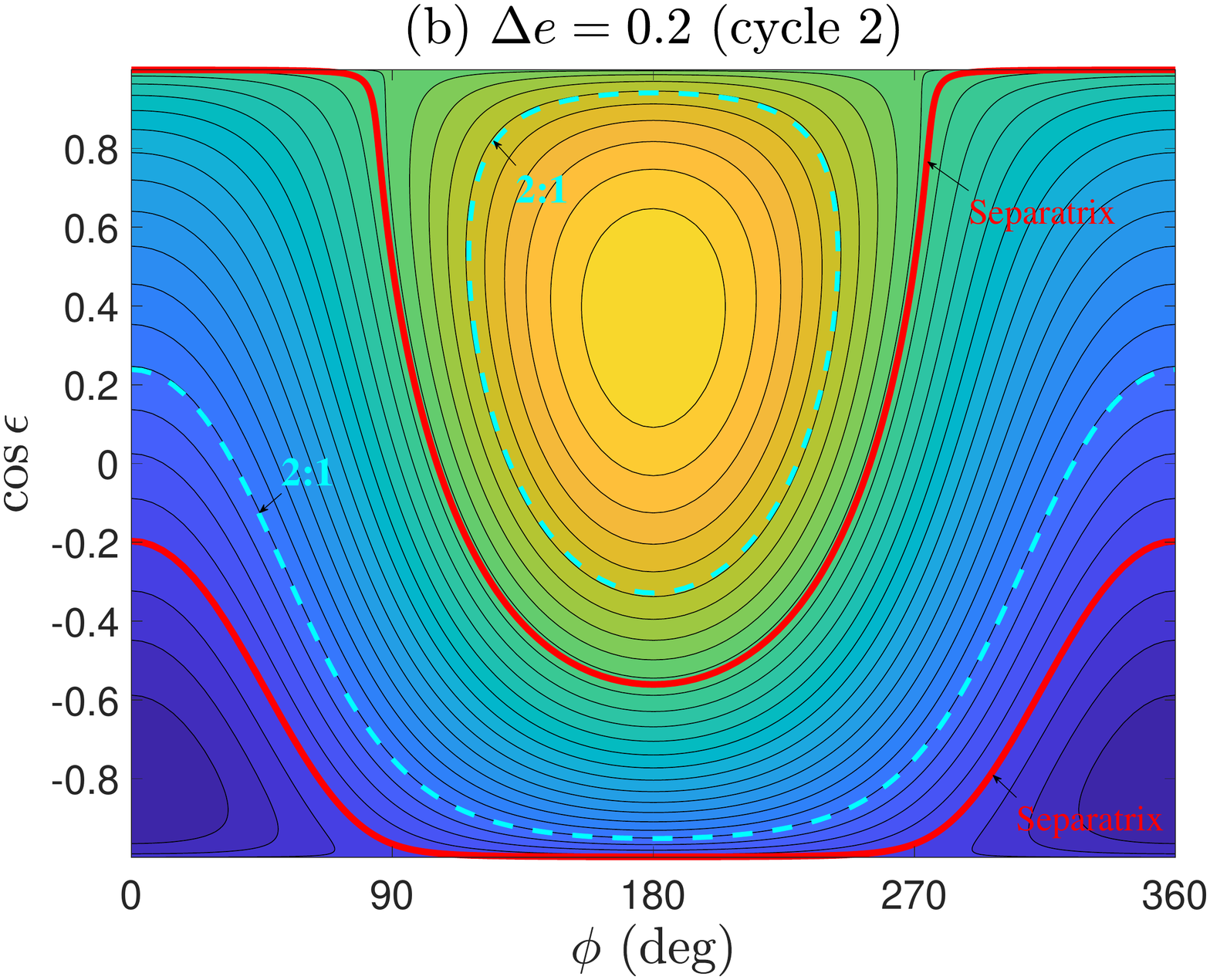}\\
	\includegraphics[width=\columnwidth]{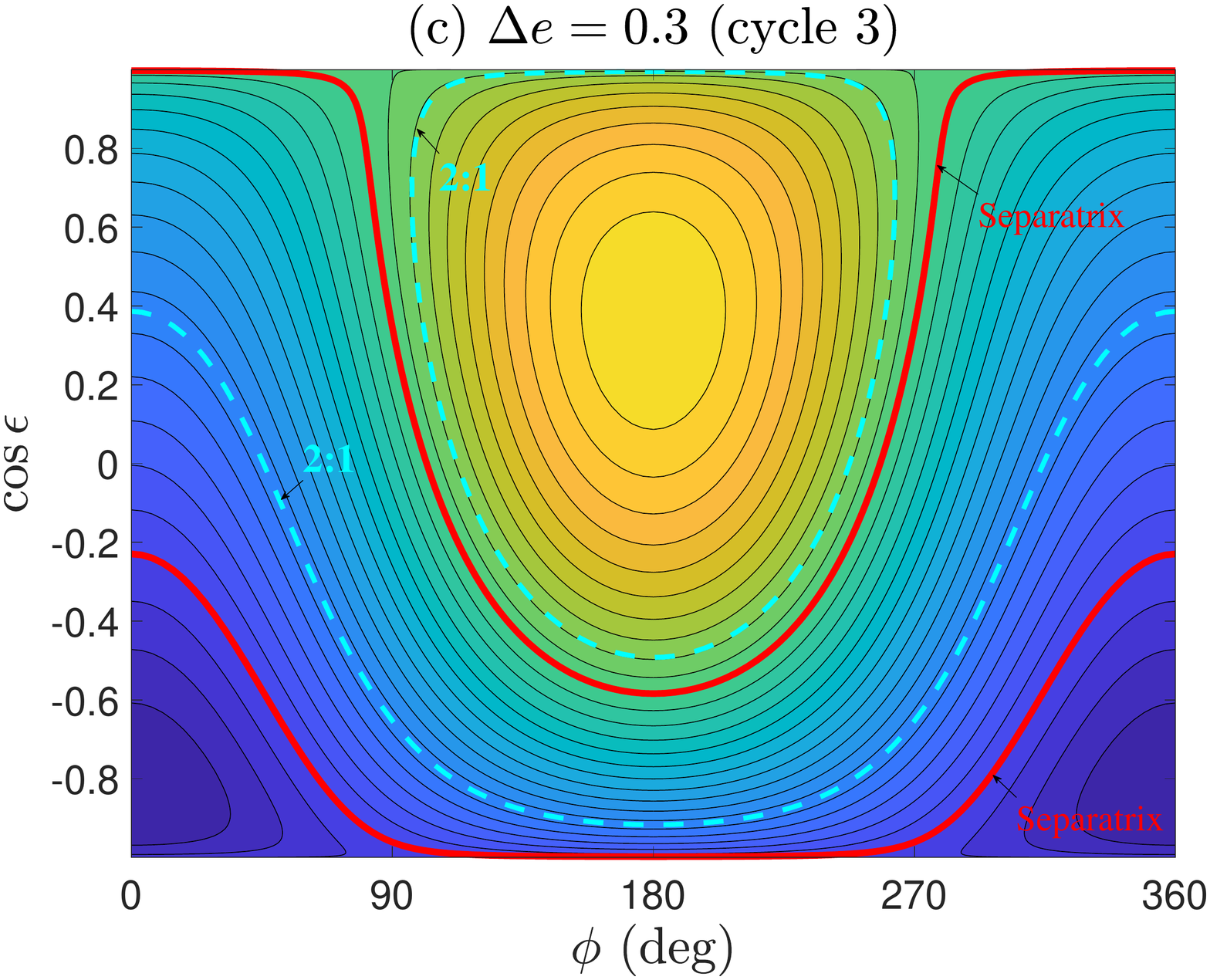}
	\includegraphics[width=\columnwidth]{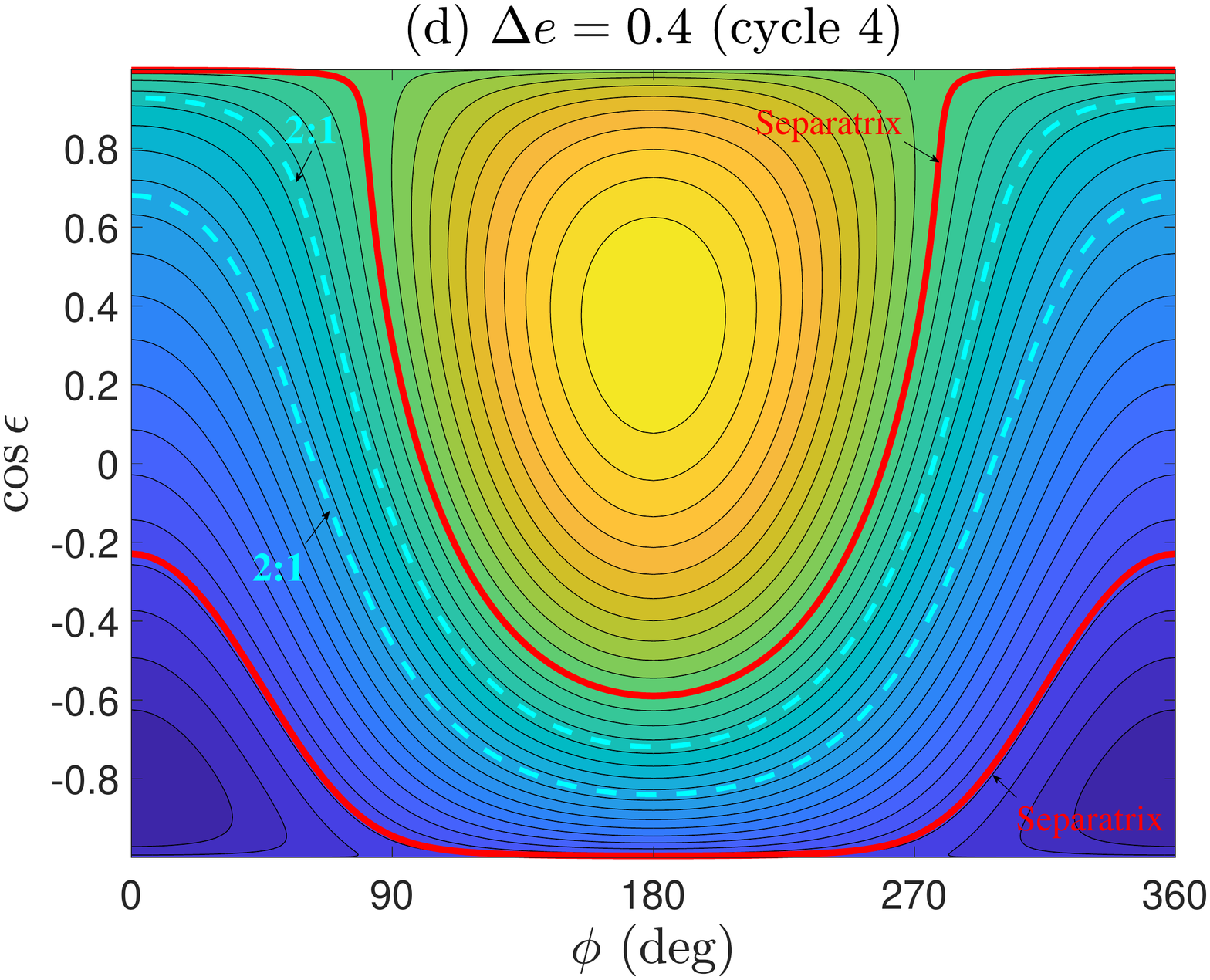}
	\caption{Dynamical structures of the unperturbed Hamiltonian model when the planet is moving on KL cycles 1, 2, 3 and 4. The red lines represent the dynamical separatrices, dividing the whole phase space into regions of circulation and of libration. The cyan dashed lines stand for the nominal locations of 2:1 spin-orbit (high-order and/or secondary) resonances. In the case of cycle 4, both 2:1 resonances occur outside the primary resonance (i.e., no 2:1 secondary resonance).}
	\label{Fig10}
\end{figure*}

According to equation (\ref{Eq16}), we can see that the Hamiltonian can be divided into two parts from the viewpoint of perturbative treatment:
\begin{equation}\label{Eq19}
	{\cal H} = {{\cal H}_0}{\rm{ + }}{{\cal H}_1},
\end{equation}
where the unperturbed Hamiltonian is
\begin{equation}\label{Eq20}
	{{\cal H}_0} = T + \frac{{\bar \alpha }}{{{n_e}}}\left( { - \frac{1}{2}{p^2} + {A_0}p + {B_0}\sqrt {1 - {p^2}} \cos \phi } \right)
\end{equation}
and the part of perturbation is given by
\begin{equation}\label{Eq21}
	\begin{aligned}
		{{\cal H}_1} =& \frac{{\bar \alpha }}{{{n_e}}}\left\{ {p\sum\limits_{n \ge 1} {{A_n}\cos \left( {n\tau } \right)}  + \frac{1}{2}\sqrt {1 - {p^2}}  \times } \right.\\
		&\left. {\sum\limits_{n \ge 1} {\left[ {\left( {{B_n} + {C_n}} \right)\cos \left( {\phi  - n\tau } \right) + \left( {{B_n} - {C_n}} \right)\cos \left( {\phi  + n\tau } \right)} \right]} } \right\}.
	\end{aligned}
\end{equation}
The unperturbed Hamiltonian (\ref{Eq20}) is equivalent to the $N=0$ Hamiltonian given in \citet{storch2017dynamics} (see equation (29) in their study).

In this section, we will discuss the spin dynamics described by the unperturbed Hamiltonian ${\cal H}_0$. Under the unperturbed Hamiltonian model, $\tau$ is absent, showing that its conjugate momentum $T$ becomes an integral of motion. Similar to the discussions made in Sect. \ref{Sect3}, the global structures in the phase space can be explored by plotting level curves of the unperturbed Hamiltonian (i.e., producing phase portraits).

In Fig. \ref{Fig10}, dynamical structures under the unperturbed Hamiltonian model are presented for the cases of KL cycles 1, 2, 3 and 4. It is observed that these four phase portraits have two islands of libration: one is centred at $\phi = \pi$ and the other one is centred at $\phi = 0$. The differences arising in the structures of these phase portraits are very small, showing that the basic dynamics governed by the unperturbed Hamiltonian is not sensitive to the amplitude of KL cycles. The dynamical separatrices are shown by red lines. Also, we can see that the configuration with $\epsilon = 0$ or $\pi$ is marginally stable because it is located close to the separatrix.

To study spin-orbit resonances, the following set of action-angle variables are introduced \citep{morbidelli2002modern}:
\begin{equation}\label{Eq22}
	{\phi ^*} = \phi  - {\rho _\phi }\left( {{\phi ^*},{p^*}} \right) = \frac{{2\pi }}{{{T_\phi }}}\tau ,\quad {p^*} = \frac{1}{{2\pi }}\oint {p{\rm d}\phi}
\end{equation}
for the case inside the island of libration, and
\begin{equation}\label{Eq23}
	{\phi ^*} = \phi  - {\rho _\phi }\left( {{\phi ^*},{p^*}} \right) = \frac{{2\pi }}{{{T_\phi }}}\tau ,\quad {p^*} = \frac{1}{{2\pi }}\int\limits_0^{2\pi } {p{\rm d}\phi }
\end{equation}
for the case outside the island of libration. $T_{\phi}$ stands for the period of $\phi$ under the unperturbed Hamiltonian model. ${\rho _\phi }$ are a periodic function of $\tau$ and its period is equal to $T_{\phi}$. $p^*$ is referred to as Arnold action \citep{morbidelli2002modern}, and $\phi^*$ is a linear function of $\tau$.

The transformations (\ref{Eq22}) and (\ref{Eq23}) are canonical with the generating function
\begin{equation*}
	S\left( {\phi ,{p^*}} \right) = \int {p\left[ {{{\cal H}_0}\left( {{p^*}} \right),\phi } \right]{\rm d}\phi }.
\end{equation*}
After the transformation, the unperturbed Hamiltonian is independent on the angle variable $\phi^*$, indicating that its conjugate momentum $p^*$ becomes an integral of motion under the unperturbed Hamiltonian model. It means the unperturbed Hamiltonian is only dependent on the action variables,
\begin{equation}\label{Eq24}
	{{\cal H}_0}\left( {T;\phi ,p} \right) \to {{\cal H}_0}\left( {T;{p^*}} \right).
\end{equation}
As a result, the fundamental frequencies can be obtained by
\begin{equation}\label{Eq25}
	\left({\phi ^*}\right)' = \frac{{\partial {{\cal H}_0}\left( {T;{p^*}} \right)}}{{\partial {p^*}}},\quad {\tau}' = \frac{{\partial {{\cal H}_0}\left( {T;{p^*}} \right)}}{{\partial T}} = 1.0.
\end{equation}
Under the unperturbed Hamiltonian model, the normalised spin frequency $\left({\phi ^*}\right)'$ changes in the interval $\left[0.446, 0.601\right]$ for KL cycle 1, in the interval $\left[0.452, 0.627\right]$ for KL cycle 2, in the interval $\left[0.465, 0.670\right]$ for KL cycle 3, and in the interval $\left[0.494, 0.741\right]$ for KL cycle 4. We know that the normalised frequency of KL cycles is equal to 1. Thus, for the considered configurations, the spin frequency is always smaller (but not much smaller) than the frequency of KL oscillation. According to the classifications discussed in \citet{storch2014chaotic}, we can get that the dynamical systems considered in this work lie in regime I (``non-adiabatic''), where the spin axis $\hat {\bm S}$ is expected to precess around the companion's orbital axis $\hat {\bm L}_b$ effectively.

It is remarked that \citet{storch2015chaotic} and \citet{storch2017dynamics} studied the dynamics of stellar spin under the configurations with planets moving on KL circulating cycles in the adiabatic and non-adiabatic regimes, respectively. In the adiabatic regime, it is shown that the dynamical structures in the time-independent Hamiltonian model are nearly symmetric with respect to $p = 0$ (see Figs. 4 and 6 in \citeauthor{storch2015chaotic} \citeyear{storch2015chaotic}). However, it is totally different for the results in the non-adiabatic regime (see Fig. 5 in \citeauthor{storch2017dynamics} \citeyear{storch2017dynamics} and Fig. \ref{Fig11} in the current work).

Spin-orbit resonance happens if the following condition is satisfied:
\begin{equation}\label{Eq26}
	k_1 \left({\phi ^*}\right)' - k_2 {\tau}' = k_1 \left({\phi ^*}\right)' - k_2 = 0,\quad k_1 \in \mathbb{N},\;k_2 \in \mathbb{Z}.
\end{equation} 
The associated critical argument of $k_1$:$k_2$ resonance is $\sigma = k_1 \phi^* - k_2 \tau$. Considering the range of spin frequency $\left({\phi ^*}\right)'$, the 2:1 spin-orbit resonance may happen in the phase space. If the 2:1 spin-orbit resonance takes place inside the primary resonance, we call it the 2:1 secondary resonance, otherwise we call it the 2:1 high-order resonance. The nominal locations of 2:1 spin-orbit resonance ($k_1 = 2, k_2 = 1$) are shown in Fig. \ref{Fig9} and Fig. \ref{Fig10} in cyan dashed lines. It is observed that both the 2:1 high-order and secondary resonances can happen in the cases of $\Delta e = 0.1, 0.2, 0.3$ while only the 2:1 high-order resonance can take place in the case of $\Delta e = 0.4$.

In the following subsection, we will further study the 2:1 high-order (or secondary) resonance in detail by taking advantage of canonical perturbation theory developed by \citet*{henrard1986perturbation} and \citet*{henrard1990semi}. This theory has been adopted to study orbital flips induced by eccentric KL mechanism in restricted hierarchical planetary systems \citep{lei2022dynamical} and in the non-restricted hierarchical planetary systems \citep{lei2022quadrupole}.

\subsection{Perturbative treatments}
\label{Sect4-4}

Through the canonical transformations shown by (\ref{Eq22}) or (\ref{Eq23}), the Hamiltonian (\ref{Eq16}) can be described as a function of the new set of variables $(\tau ,{\phi ^*},T,{p^*})$ by
\begin{equation}\label{Eq27}
	{\cal H}\left( {\tau ,{\phi ^*},T,{p^*}} \right) = {\cal H}_0\left( {T,{p^*}} \right) + {\cal H}_1\left( {\tau ,{\phi ^*},T,{p^*}} \right).
\end{equation}
It is noted that it is very difficult to provide the explicit expression of ${\cal H}$ for equation (\ref{Eq27}) because the transformation between $(\phi, p)$ and $(\phi^*, p^*)$ given by equation (\ref{Eq22}) or (\ref{Eq23}) is achieved by means of numerical quadrature.

Usually, ${\cal H}_1$ is much smaller than ${\cal H}_0$. From the viewpoint of perturbative treatment, the term ${\cal H}_0$ determines the unperturbed dynamical model (or Kernel Hamiltonian model) and the term ${\cal H}_1$ plays the role of perturbation to the unperturbed Hamiltonian model. \citet{storch2017dynamics} studied the spin dynamics in the non-adiabatic and adiabatic regimes under the unperturbed Hamiltonian model specified by ${\cal H}_0$. However, a more formal canonical perturbation theory considering the contribution of the higher-order Hamiltonian is absent in their work.

To study the 2:1 spin-orbit resonance, let us introduce the following linear transformation:
\begin{equation}\label{Eq28}
	\begin{aligned}
		{\sigma _1} &= {\phi ^*} - \frac{1}{2}\tau ,\quad {\Sigma _1} = {p^*},\\
		{\sigma _2} &= \tau ,\quad\quad {\Sigma _2} = T + \frac{1}{2}{p^*}.
	\end{aligned}
\end{equation}
Here, $\sigma_1$ is the resonant argument of the 2:1 spin-orbit resonance. The transformation (\ref{Eq28}) is canonical with the generating function,
\begin{equation*}
	S\left( {\tau ,{\phi ^*},{\Sigma _1},{\Sigma _2}} \right) = {\phi ^*}{\Sigma _1} + \tau \left( {{\Sigma _2} - \frac{1}{2}{\Sigma _1}} \right).
\end{equation*}
Using the transformation (\ref{Eq28}), the Hamiltonian (\ref{Eq27}) can be organised as
\begin{equation}\label{Eq29}
	{\cal H}\left( {\sigma_1 ,\sigma_2,\Sigma_1,\Sigma_2} \right) = {\cal H}_0\left( {\Sigma_1,\Sigma_2} \right) + {\cal H}_1\left( {\sigma_1 ,\sigma_2,\Sigma_1,\Sigma_2} \right),
\end{equation}
and the equations of motion can be obtained by Hamiltonian canonical relations:
\begin{equation}\label{Eq30}
	\begin{aligned}
		\frac{{{\rm d}{\sigma _1}}}{{{\rm d}\tau}} &= \frac{{\partial {\cal H}}}{{\partial {\Sigma _1}}},\quad \frac{{{\rm d}{\Sigma _1}}}{{{\rm d}\tau}} =  - \frac{{\partial {\cal H}}}{{\partial {\sigma _1}}},\\
		\frac{{{\rm d}{\sigma _2}}}{{{\rm d}\tau}}&= \frac{{\partial {\cal H}}}{{\partial {\Sigma _2}}},\quad \frac{{{\rm d}{\Sigma _2}}}{{{\rm d}\tau}} =  - \frac{{\partial {\cal H}}}{{\partial {\sigma _2}}}.
	\end{aligned}
\end{equation}

In particular, when the configuration is inside the 2:1 spin-orbit resonance, the critical argument $\sigma_1$ becomes a long-period variable, and the angle $\sigma_2$ is a fast angular variable. Thus, this is a typical separable Hamiltonian model. According to mean element theory \citep{kozai1959motion}, the terms in Hamiltonian (\ref{Eq29}) can be classified into secular, long-period and short-period terms. The long-term evolution of stellar spin is dominated by the secular and long-period terms in the Hamiltonian. As a result, those short-period effects can be removed.

In practice, we adopt averaging technique (the lowest-order perturbation theory) to filter out those short-period effects in order to formulate the resonant Hamiltonian as follows: 
\begin{equation}\label{Eq31}
	\begin{aligned}
		{{\cal H}^*} =&\; \frac{1}{{4\pi }}\int\limits_0^{4\pi } {{\cal H}\left( {{\sigma _1},{\sigma _2},{\Sigma _1},{\Sigma _2}} \right){\rm d}{\sigma _2}}\\
		=&\; {\cal H}_0 \left( \Sigma_1, \Sigma_2 \right) + \frac{1}{{4\pi }}\int\limits_0^{4\pi } {{\cal H}_1\left( {{\sigma _1},{\sigma _2},{\Sigma _1},{\Sigma _2}} \right){\rm d}{\sigma _2}}.
	\end{aligned}
\end{equation}
The angular variable $\sigma_2$ is absent from the resonant Hamiltonian ${{\cal H}^*}$, thus its conjugate momentum $\Sigma_2$ becomes the motion integral under the resonant model, leading to the fact that the resonant model is of one degree of freedom.

\begin{figure*}
	\centering
	\includegraphics[width=0.9\columnwidth]{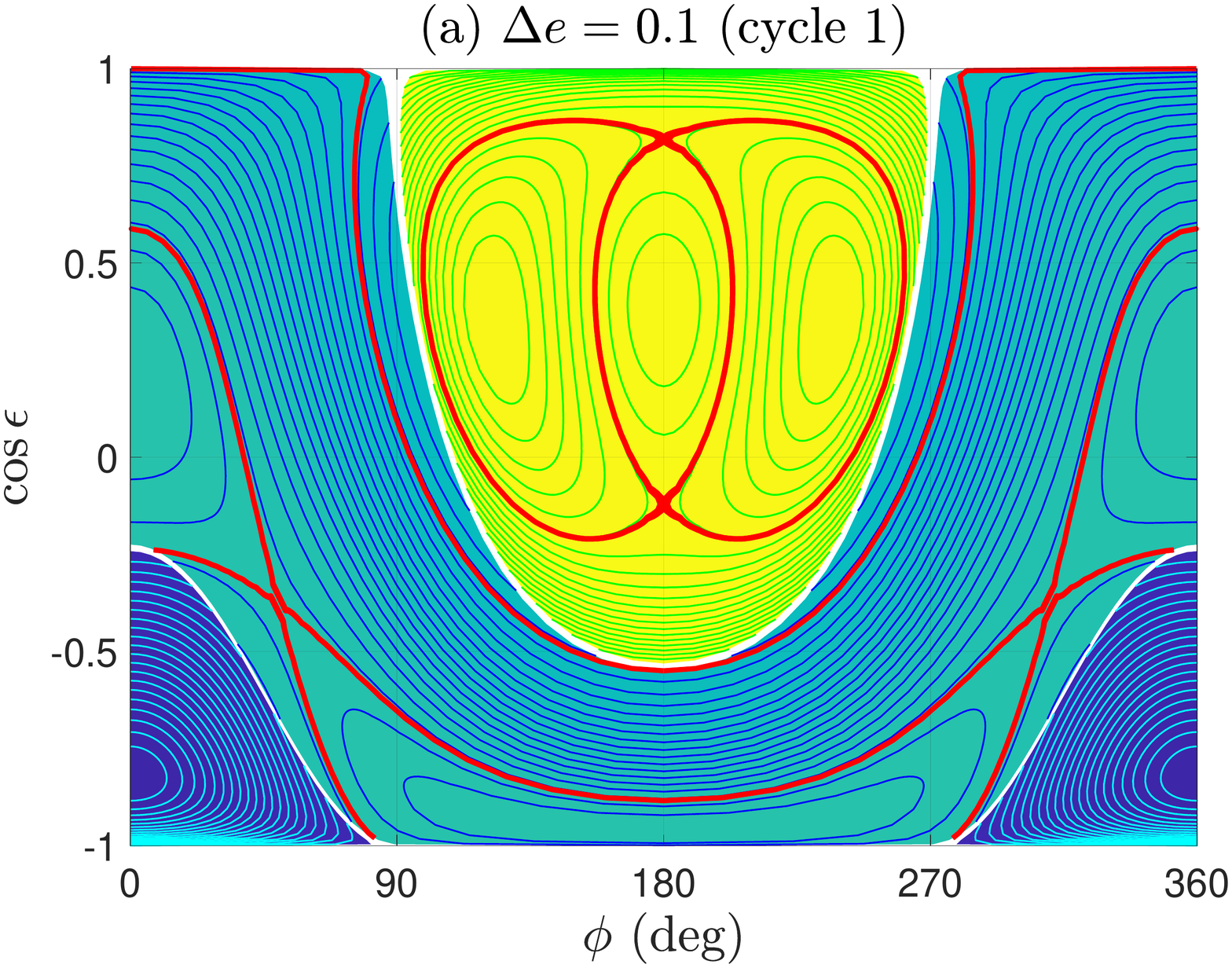}
	\includegraphics[width=0.9\columnwidth]{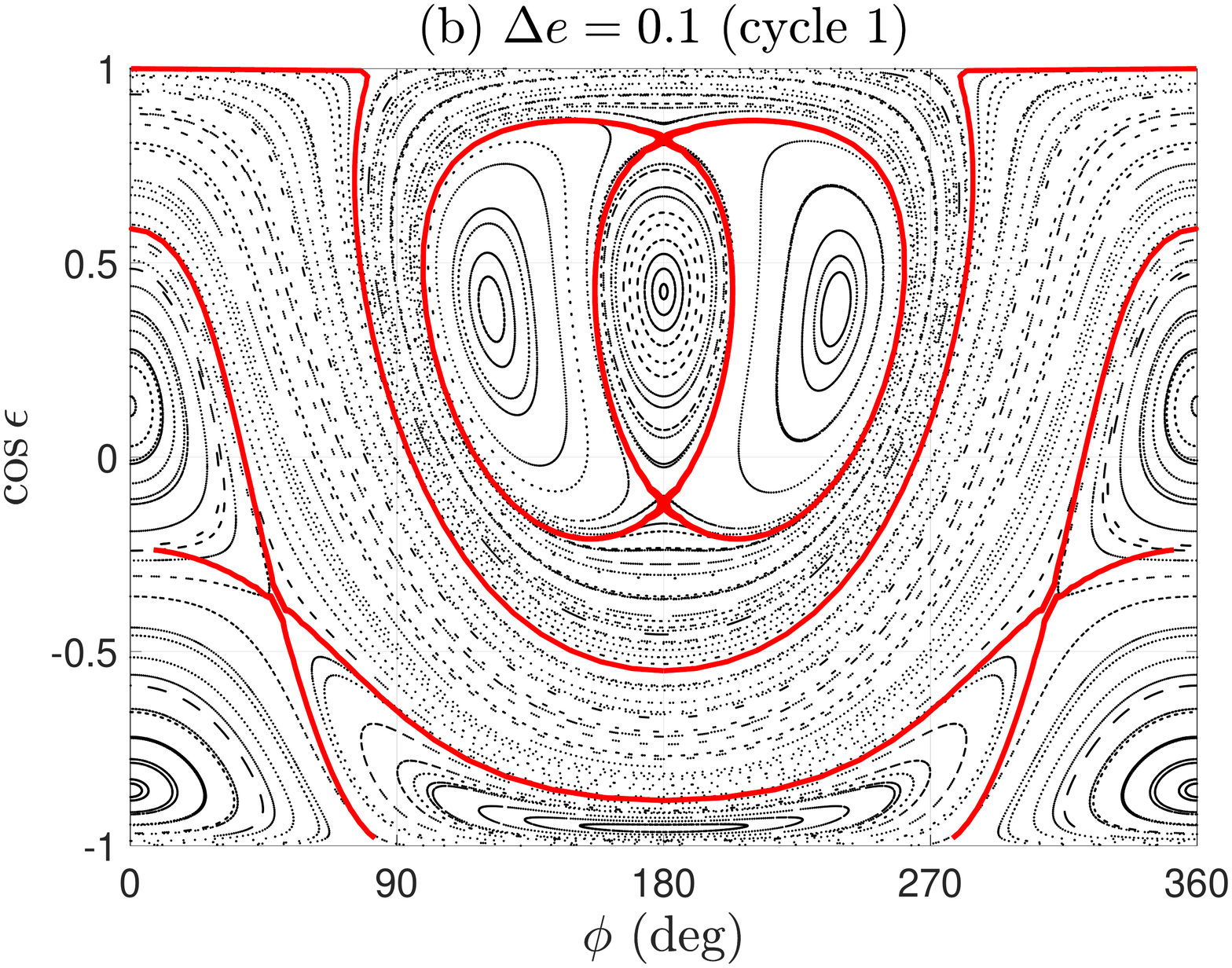}\\
	\includegraphics[width=0.9\columnwidth]{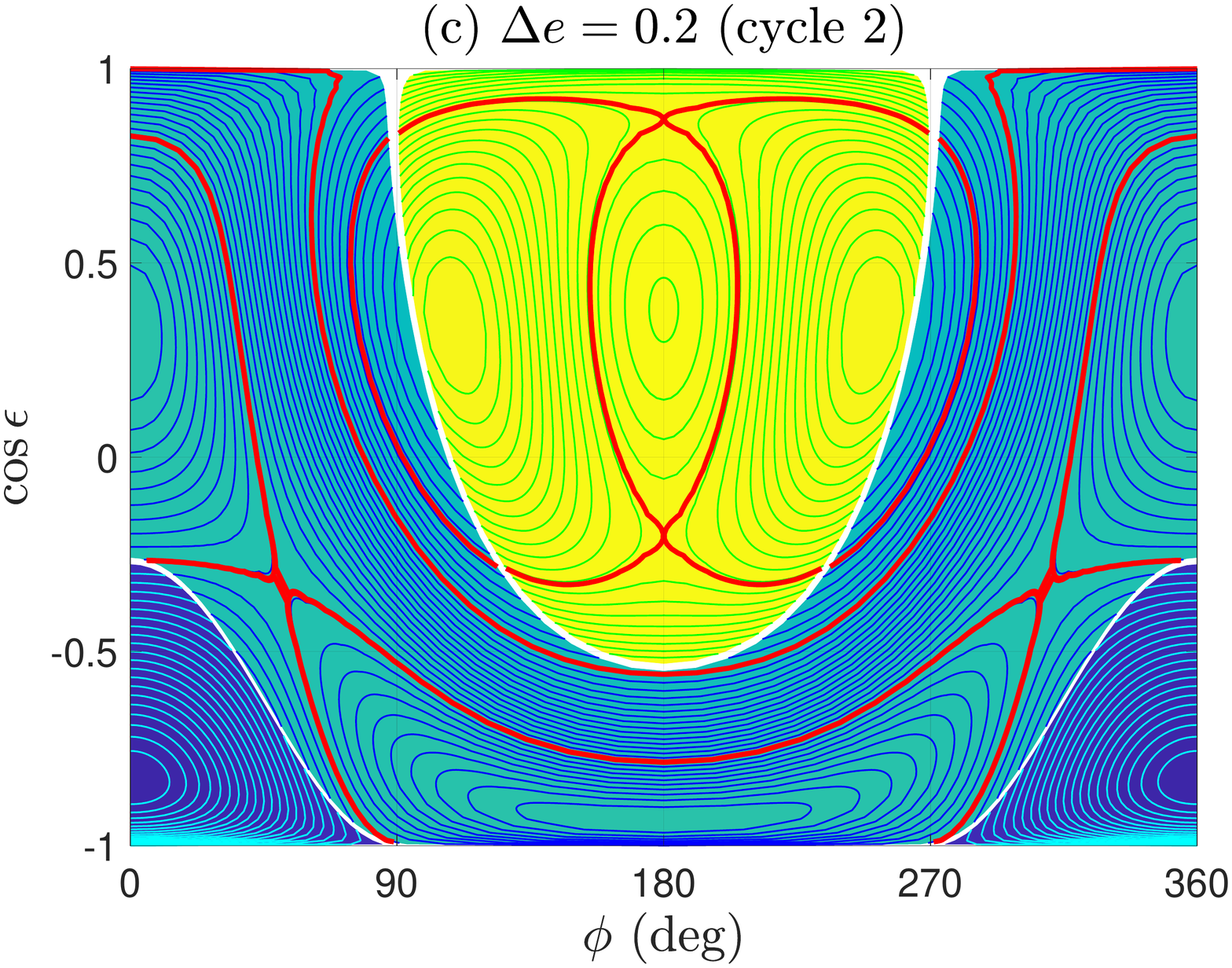}
	\includegraphics[width=0.9\columnwidth]{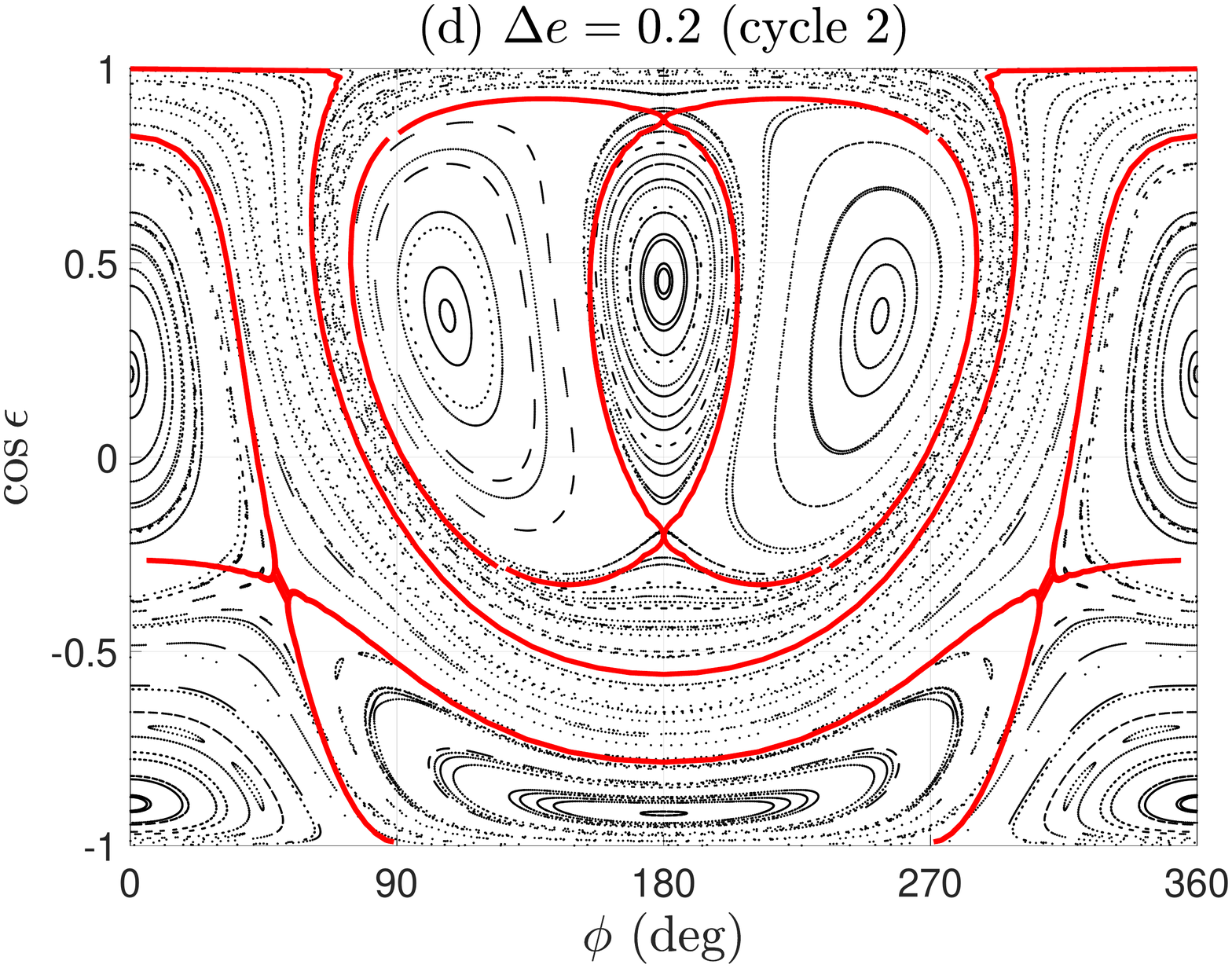}\\
	\includegraphics[width=0.9\columnwidth]{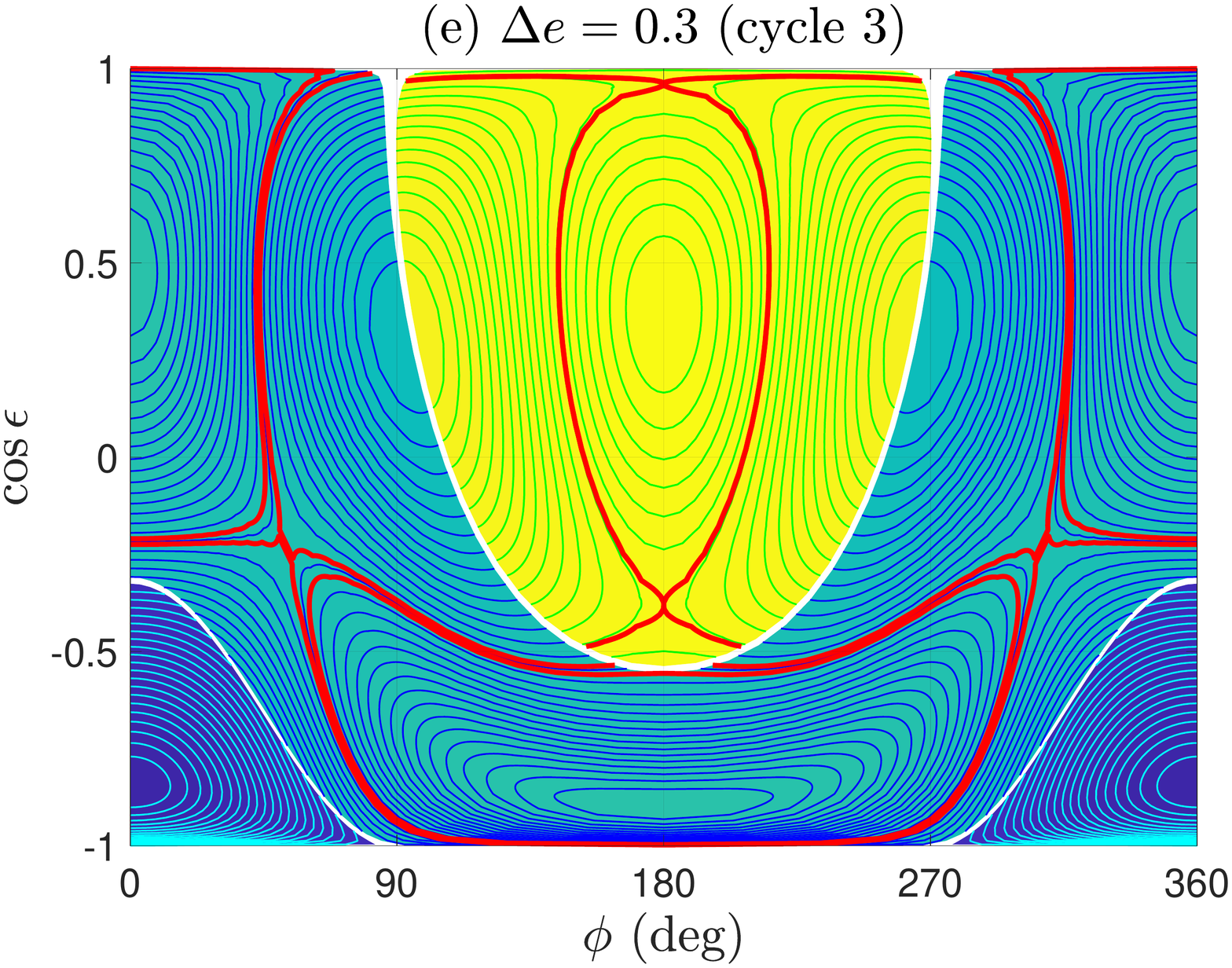}
	\includegraphics[width=0.9\columnwidth]{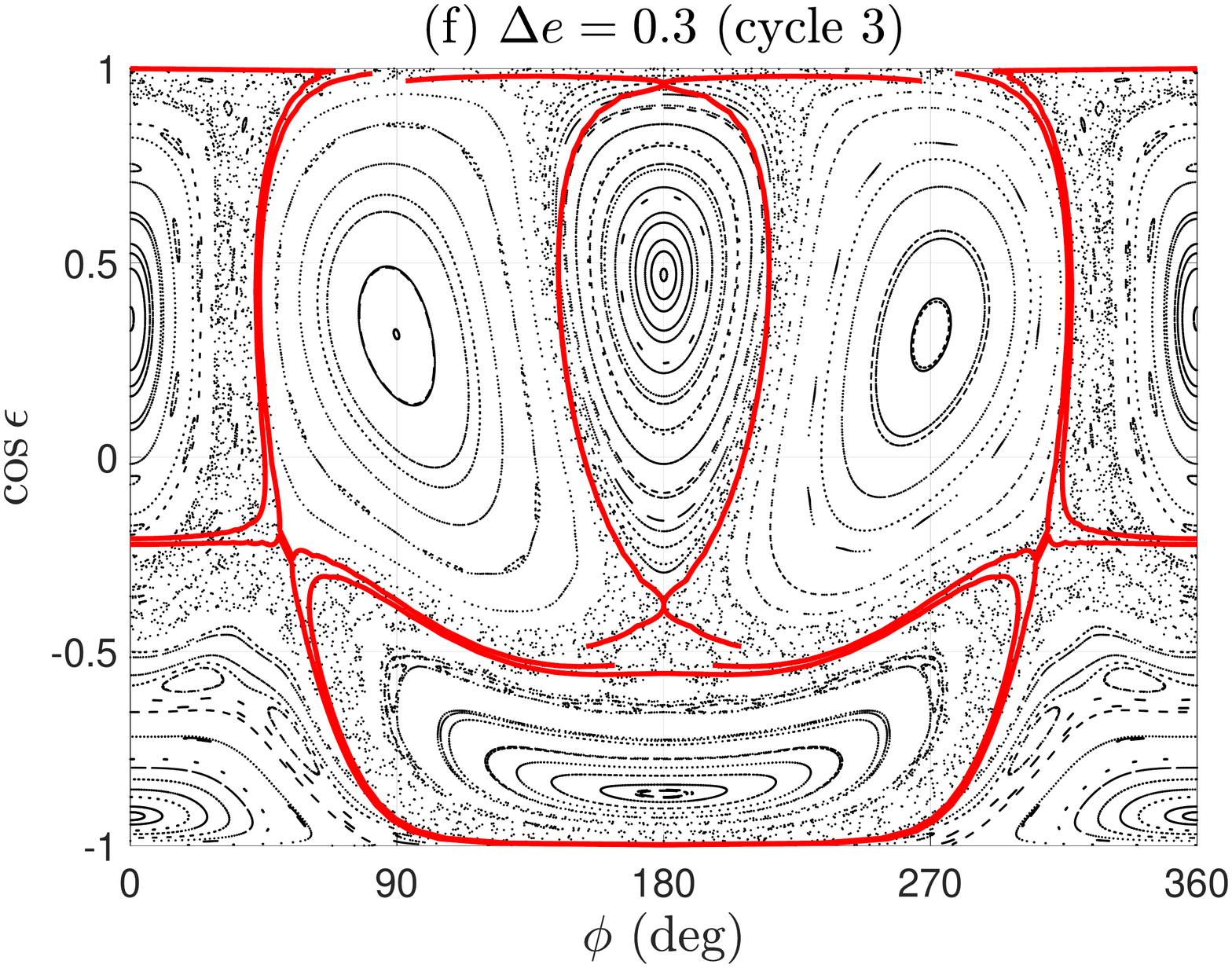}\\
	\includegraphics[width=0.9\columnwidth]{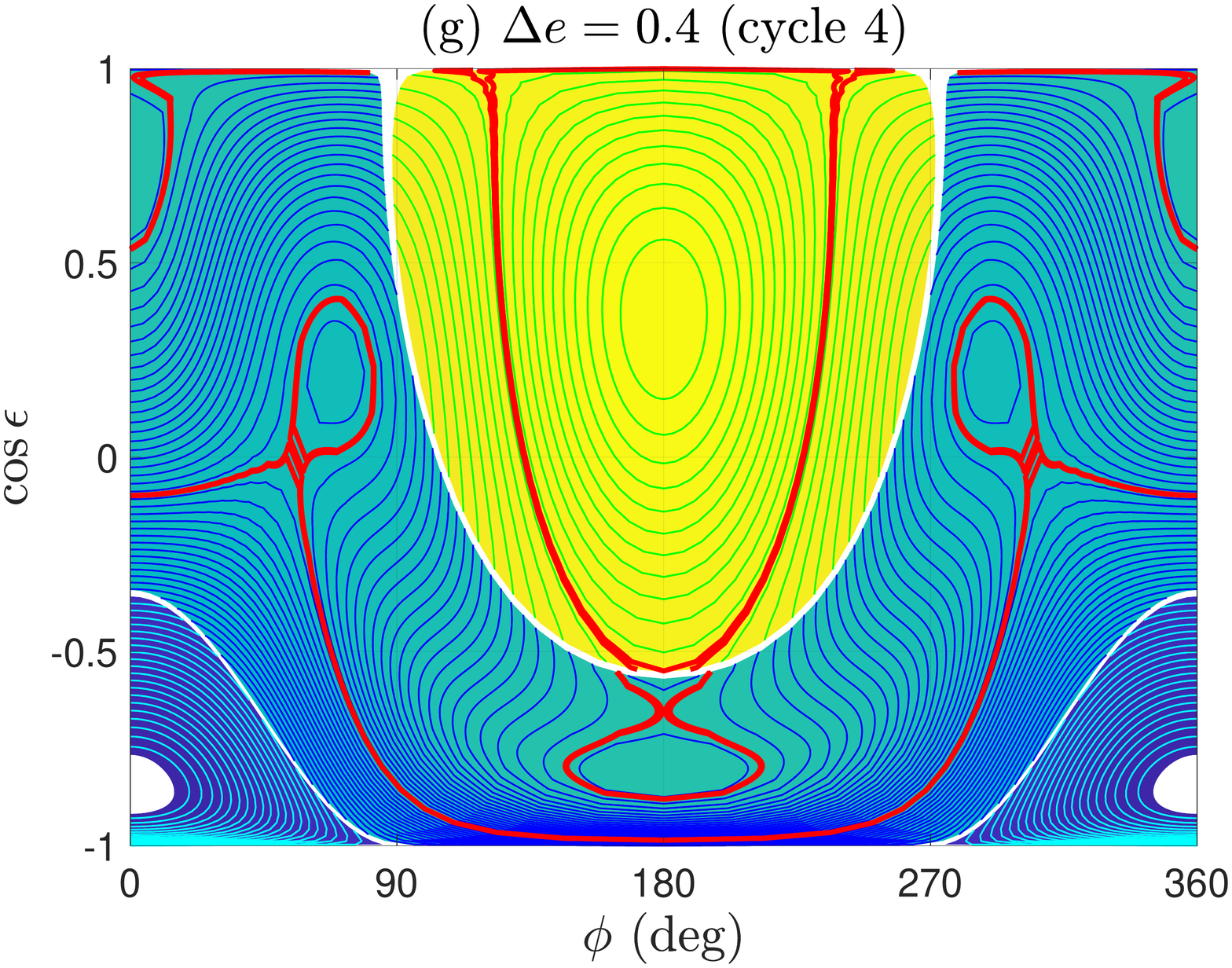}
	\includegraphics[width=0.9\columnwidth]{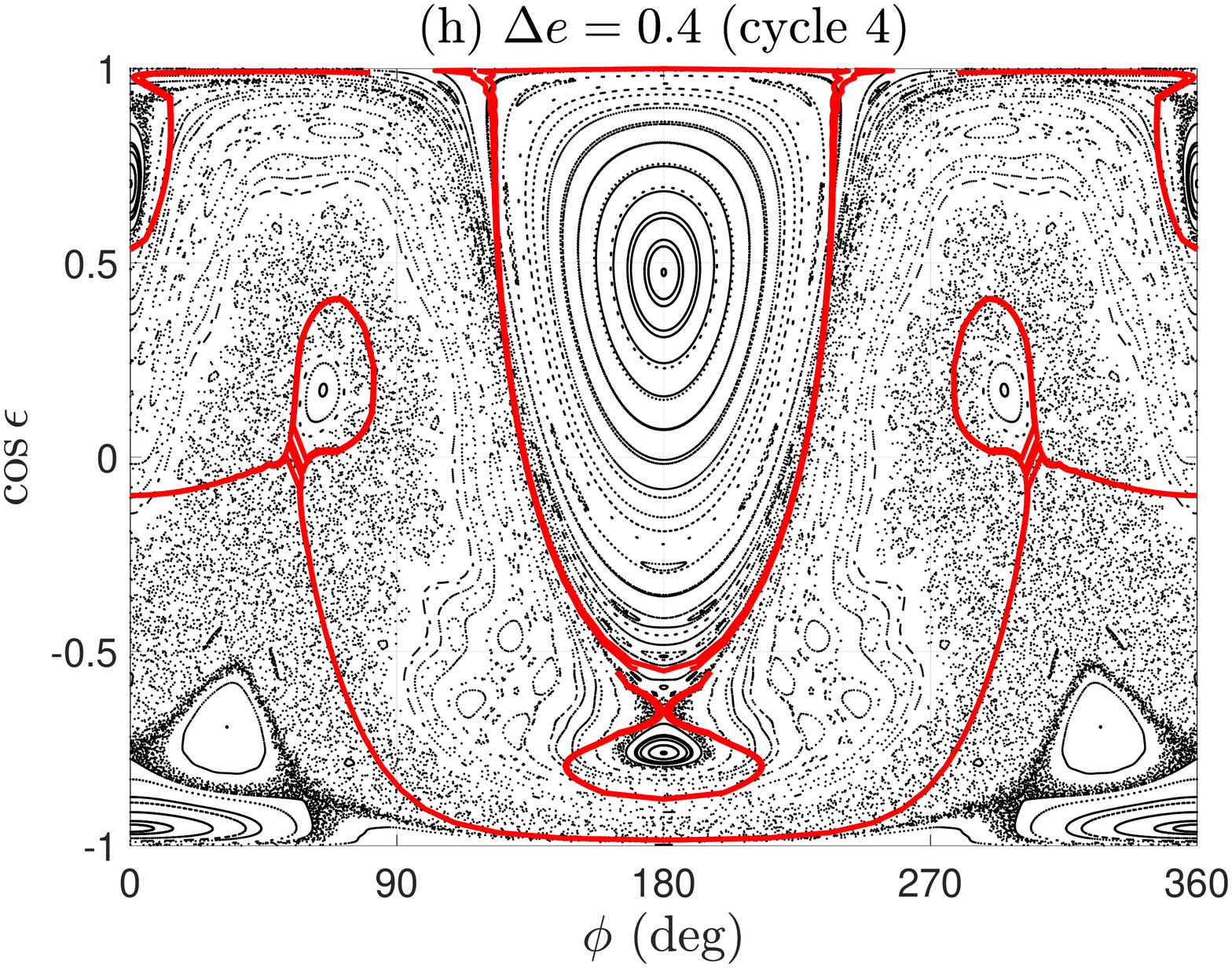}
	\caption{Analytical structures in the phase portraits under the resonant model (\textit{left-column panels}) and the related numerical structures in the Poincar\'e sections (\textit{right-column panels}) for secular dynamics of stellar spin under the configurations with planets moving on KL cycles 1--4. The red lines in all panels are dynamical separatrices, which correspond to the level curves of resonant Hamiltonian passing through saddle points.}
	\label{Fig11}
\end{figure*}

Under the resonant model, the dynamical structures can be explored by taking advantage of phase portraits (i.e., level curves of resonant Hamiltonian). When the motion integral $\Sigma_2 = T + \frac{1}{2}{p^*}$ is given, the dynamical structures in the $(\sigma_1,\Sigma_1)$ plane can be obtained. For convenience, we project phase portraits shown in the $(\sigma_1,\Sigma_1)$ plane to the ones in the $(\phi, \cos{\epsilon})$ plane. Here $(\phi, \cos{\epsilon})$ corresponds to the spin state of star when the eccentricity of planet reaches a maximum, corresponding to the points on Poincar\'e section defined by equation (\ref{Eq19}). By doing so can we compare the phase portraits and Poincar\'e surfaces of section directly.

The main results of this work are presented in Fig. \ref{Fig11} for stellar spin dynamics under the configurations with planets moving on KL cycles 1--4. In the left panels, phase portraits under the resonant model are provided and, in the right panels, the associated Poincar\'e sections are shown. It can be observed that analytical structures in the phase portraits agree well with the numerical structures in the Poincar\'e sections, supporting the conclusion that the complex dynamics of stellar spin are governed by the primary resonance under the unperturbed Hamiltonian model in combination with the 2:1 (high-order and/or secondary) spin-orbit resonances.

The red lines shown in the left panels of Fig. \ref{Fig11} stand for the dynamical separatrices, which divide the entire phase space into regions of libration and circulation. For convenience of comparison, the dynamical separatrices are also shown on the Poincar\'e sections. It is observed that the dynamical separatrices can provide good boundaries for islands of libration in the Poincar\'e sections. In addition, chaotic layers are distributed around the dynamical separatrices. 

It is remarked that, for the cases of cycles 3 and 4, libration islands caused by higher-order spin-orbit resonances can be found. In addition, wider chaotic layers can be observed around the dynamical separatrices due to stronger perturbation (compared to the cases of cycles 1 and 2). However, the main structures arising in the Poincar\'e sections can be well understood with the aid of resonant dynamics.

\section{Conclusions}
\label{Sect5}

In this study, secular dynamics of stellar spin are investigated by means of numerical and analytical approaches under the hierarchical configuration where the planet is undergoing KL libration driven by a distant and inclined binary companion. About the dynamical model, it is assumed that the orbital evolution of the bodies involved is independent upon the stellar rotation (i.e., back-reaction effect is ignored). The planet is assumed to move on KL librating cycles with zero and nonzero amplitudes (it is noted that KL cycle with zero amplitude corresponds to the KL fixed point). The zero-amplitude case can be used to approximate those dynamical configurations which are deeply inside KL resonance. The nonzero-amplitude cases can approximate more general configurations with planets moving on KL librating cycles. 

When the planet is assumed at the KL fixed point, the Hamiltonian governing stellar spin determines a one-degree-of-freedom dynamical system.  The phase space structures are revealed by plotting level curves of the Hamiltonian. In particular, when the maximum inclination $i_{\max}$ is smaller than $\sim$$80^{\circ}$, phase portraits of stellar spin exhibit similar structures: there are two islands of libration, one is centred at $\phi = \pi$ with $\epsilon < 90^{\circ}$ and the other one is centred at $\phi = 0$ with $\epsilon > 90^{\circ}$. However, when the maximum inclination $i_{\max}$ is greater than $\sim$$80^{\circ}$, besides the islands of libration arising in the former case, an additional island of libration centred at $\phi = 0$ with $\epsilon < 90^{\circ}$ appears. It is found that the resonant width of stellar spin first increases and then decreases with $i_{\max}$ for all the three branches of libration.

When the planet is assumed on nonzero-amplitude KL cycles, the approximate Hamiltonian determines a 1.5-degree-of-freedom dynamical model. In order to study spin-orbit resonances, the Hamiltonian is further augmented into a two-degree-of-freedom system by introducing an action conjugated to the time-like variable $\tau$. For such a 2-DOF Hamiltonian model, the technique of Poincar\'e sections is applied and the global structures in the phase space are revealed. There are complex structures arising in Poincar\'e sections: regular and chaotic behaviours of stellar spin can be observed. In order to understand the basic structures, the perturbative treatments developed by \citet{henrard1986perturbation} are adopted. In particular, the spin Hamiltonian is divided into two parts: the unperturbed Hamiltonian and the part of perturbation. In terms of magnitude, the part of perturbation is much smaller than that of the unperturbed one. The unperturbed Hamiltonian specifies an integrable dynamical model, where the global dynamical structures can be explored by means of phase portraits. By analysing the distribution of unperturbed fundamental frequencies, it is found that the 2:1 high-order and/or secondary spin-orbit resonances happen in the phase space. To study the 2:1 spin--orbit resonances, we introduce a canonical transformation, which makes the Hamiltonian model be separable in terms of fundamental frequencies. Thus, it is possible to formulate the resonant Hamiltonian by performing an average for the Hamiltonian over the period of the fast degree of freedom. The resulting resonant model is a one-degree-of-freedom Hamiltonian system, where the global dynamical structures can be explored by analysing phase portraits.

Our main results are provided in Fig. \ref{Fig11}. It is found that analytical structures in phase portraits can agree well with the numerical structures arising in Poincar\'e sections, indicating that the complex dynamics of stellar spin in the phase space are dominated by the primary resonance under the unperturbed Hamiltonian model in combination with the 2:1 (high-order and/or secondary) spin-orbit resonances. The dynamical separatrices determined under the resonant model can provide good boundaries for islands of libration. Chaotic layers are distributed around dynamical separatrices and they become wider when the amplitude of KL cycle is larger. At last, it is observed that the configurations with $\epsilon = 0$ and $\epsilon = \pi$ are marginally stable because they are close to the separatrices.

\section*{Acknowledgements}
Hanlun Lei wishes to thank an anonymous reviewer for helpful suggestions that improved the quality of this manuscript. This work is supported by the National Natural Science Foundation of China (Nos. 12073011, 12073019 and 12233003) and the National Key R\&D Program of China (No. 2019YFA0706601).

\section*{Data availability}
The analysis and codes are available upon request.

\bibliographystyle{mn2e}
\bibliography{mybib}

\begin{thebibliography}{}

\bibitem[\protect\citeauthoryear{Albrecht, Dawson \& Winn}{Albrecht
  et~al.}{2022}]{albrecht2022stellar}
Albrecht S.~H.,  Dawson R.~I.,    Winn J.~N.,  2022, PASP, 134, 082001

\bibitem[\protect\citeauthoryear{Anderson, Lai \& Storch}{Anderson
  et~al.}{2017}]{anderson2017eccentricity}
Anderson K.~R.,  Lai D.,    Storch N.~I.,  2017, MNRAS, 467, 3066

\bibitem[\protect\citeauthoryear{Antognini}{Antognini}{2015}]{antognini2015timescales}
Antognini J.~M.,  2015, MNRAS, 452, 3610

\bibitem[\protect\citeauthoryear{Bate, Lodato \& Pringle}{Bate
  et~al.}{2010}]{bate2010chaotic}
Bate M.,  Lodato G.,    Pringle J.,  2010, MNRAS, 401, 1505

\bibitem[\protect\citeauthoryear{Batygin}{Batygin}{2012}]{batygin2012primordial}
Batygin K.,  2012, Nature, 491, 418

\bibitem[\protect\citeauthoryear{Dawson \& Johnson}{Dawson \&
  Johnson}{2018}]{dawson2018origins}
Dawson R.~I.,  Johnson J.~A.,  2018, ARA\&A, 56, 175

\bibitem[\protect\citeauthoryear{Hamers}{Hamers}{2021}]{hamers2021properties}
Hamers A.~S.,  2021, MNRAS, 500, 3481

\bibitem[\protect\citeauthoryear{Henrard}{Henrard}{1990}]{henrard1990semi}
Henrard J.,  1990, CeMDA, 49, 43

\bibitem[\protect\citeauthoryear{Henrard \& Lemaitre}{Henrard \&
  Lemaitre}{1986}]{henrard1986perturbation}
Henrard J.,  Lemaitre A.,  1986, Celest. Mech., 39, 213

\bibitem[\protect\citeauthoryear{Huang \& Lei}{Huang \&
  Lei}{2022}]{huang2022orbital}
Huang X.,  Lei H.,  2022, AJ, 164, 232

\bibitem[\protect\citeauthoryear{Katz, Dong \& Malhotra}{Katz
  et~al.}{2011}]{katz2011long}
Katz B.,  Dong S.,    Malhotra R.,  2011, Phys. Rev. Lett., 107, 181101

\bibitem[\protect\citeauthoryear{Kinoshita}{Kinoshita}{1993}]{kinoshita1993motion}
Kinoshita H.,  1993, CeMDA, 57, 359

\bibitem[\protect\citeauthoryear{Kinoshita \& Nakai}{Kinoshita \&
  Nakai}{2007}]{kinoshita2007general}
Kinoshita H.,  Nakai H.,  2007, CeMDA, 98, 67

\bibitem[\protect\citeauthoryear{Kozai}{Kozai}{1959}]{kozai1959motion}
Kozai Y.,  1959, AJ, 64, 367

\bibitem[\protect\citeauthoryear{Kozai}{Kozai}{1962}]{kozai1962secular}
Kozai Y.,  1962, AJ, 67, 591

\bibitem[\protect\citeauthoryear{Laskar \& Robutel}{Laskar \&
  Robutel}{1993}]{laskar1993chaotic}
Laskar J.,  Robutel P.,  1993, Nature, 361, 608

\bibitem[\protect\citeauthoryear{Lei}{Lei}{2022}]{lei2022systematic}
Lei H.,  2022, AJ, 163, 214

\bibitem[\protect\citeauthoryear{Lei \& Gong}{Lei \&
  Gong}{2022}]{lei2022dynamical}
Lei H.,  Gong Y.-X.,  2022, A\&A, 665, A62

\bibitem[\protect\citeauthoryear{Lei \& Huang}{Lei \&
  Huang}{2022}]{lei2022quadrupole}
Lei H.,  Huang X.,  2022, MNRAS, 515, 1086

\bibitem[\protect\citeauthoryear{Li, Naoz, Holman \& Loeb}{Li
  et~al.}{014a}]{li2014}
Li G.,  Naoz S.,  Holman M.,    Loeb A.,  2014a, ApJ, 791, 86

\bibitem[\protect\citeauthoryear{Li, Naoz, Kocsis \& Loeb}{Li
  et~al.}{014b}]{li2014eccentricity}
Li G.,  Naoz S.,  Kocsis B.,    Loeb A.,  2014b, ApJ, 785, 116

\bibitem[\protect\citeauthoryear{Lithwick \& Naoz}{Lithwick \&
  Naoz}{2011}]{lithwick2011eccentric}
Lithwick Y.,  Naoz S.,  2011, ApJ, 742, 94

\bibitem[\protect\citeauthoryear{Liu, Munoz \& Lai}{Liu
  et~al.}{2015}]{liu2015suppression}
Liu B.,  Munoz D.~J.,    Lai D.,  2015, MNRAS, 447, 747

\bibitem[\protect\citeauthoryear{Morbidelli}{Morbidelli}{2002}]{morbidelli2002modern}
Morbidelli A.,  2002, Modern celestial mechanics: aspects of solar system
  dynamics.
Taylor \& Francis, London and New York

\bibitem[\protect\citeauthoryear{Murray \& Dermott}{Murray \&
  Dermott}{1999}]{murray1999solar}
Murray C.~D.,  Dermott S.~F.,  1999, Solar system dynamics.
Cambridge university press

\bibitem[\protect\citeauthoryear{Naoz}{Naoz}{2016}]{naoz2016eccentric}
Naoz S.,  2016, ARA\&A, 54, 441

\bibitem[\protect\citeauthoryear{Naoz, Farr, Lithwick, Rasio \&
  Teyssandier}{Naoz et~al.}{2011}]{naoz2011hot}
Naoz S.,  Farr W.~M.,  Lithwick Y.,  Rasio F.~A.,    Teyssandier J.,  2011,
  Nature, 473, 187

\bibitem[\protect\citeauthoryear{Peale}{Peale}{1969}]{peale1969generalized}
Peale S.~J.,  1969, AJ, 74, 483

\bibitem[\protect\citeauthoryear{Petrovich}{Petrovich}{2015}]{petrovich2015hot}
Petrovich C.,  2015, ApJ, 805, 75

\bibitem[\protect\citeauthoryear{Shevchenko}{Shevchenko}{2016}]{shevchenko2016lidov}
Shevchenko I.~I.,  2016, The Lidov-Kozai effect-applications in exoplanet
  research and dynamical astronomy.
Vol.~441, Springer

\bibitem[\protect\citeauthoryear{Sidorenko}{Sidorenko}{2018}]{sidorenko2018eccentric}
Sidorenko V.~V.,  2018, Celest. Mech. Dyn. Astron., 130, 4

\bibitem[\protect\citeauthoryear{Storch, Anderson \& Lai}{Storch
  et~al.}{2014}]{storch2014chaotic}
Storch N.~I.,  Anderson K.~R.,    Lai D.,  2014, Science, 345, 1317

\bibitem[\protect\citeauthoryear{Storch \& Lai}{Storch \&
  Lai}{2015}]{storch2015chaotic}
Storch N.~I.,  Lai D.,  2015, MNRAS, 448, 1821

\bibitem[\protect\citeauthoryear{Storch, Lai \& Anderson}{Storch
  et~al.}{2017}]{storch2017dynamics}
Storch N.~I.,  Lai D.,    Anderson K.~R.,  2017, MNRAS, 465, 3927

\bibitem[\protect\citeauthoryear{Wisdom, Peale \& Mignard}{Wisdom
  et~al.}{1984}]{wisdom1984chaotic}
Wisdom J.,  Peale S.~J.,    Mignard F.,  1984, Icarus, 58, 137

\bibitem[\protect\citeauthoryear{Wu \& Murray}{Wu \&
  Murray}{2003}]{wu2003planet}
Wu Y.,  Murray N.,  2003, ApJ, 589, 605

\end{thebibliography}


\bsp
\label{lastpage}
\end{document}